\documentclass[%
reprint,
superscriptaddress,
bibnotes,
amsmath,amssymb,
aps,
pra,
]{revtex4-2}

\usepackage{graphicx}% Include figure files
\usepackage[caption=false]{subfig}
\usepackage{qcircuit}
\usepackage{dcolumn}% Align table columns on decimal point
\usepackage{bm}% bold math
\usepackage{tikz}
\usetikzlibrary{plotmarks}
\usetikzlibrary{decorations.markings}
\usepackage{siunitx}
\usepackage{xy}
\usepackage{dsfont}
\usepackage{gensymb}
\usepackage{pgfplots}
\usepackage{pgf}
\pgfplotsset{compat=1.15}
\usepackage{placeins}
\usepackage{flafter}
\usepackage{color}
\usepackage{orcidlink}
\usepackage{hyperref}% add hypertext capabilities
\hypersetup{
	colorlinks=true,
	linkcolor=magenta,
	filecolor=blue,      
	urlcolor=magenta,
	citecolor=blue,
	pdftex,
	pdfauthor={Ahmed M. Farouk, I. I. Beterov, Peng Xu and I.I. Ryabtsev},
	pdftitle={Generation of quantum phases of matter and finding a maximum-weight independent set of unit-disk graphs using Rydberg atoms},
   pdfcreator={Ahmed M. Farouk},
}
\usepackage{epstopdf}
\usepackage{times}
\usepackage[varg]{txfonts}
\usepackage{textcomp}
\makeatletter
\newcommand\tinyv{\@setfontsize\tinyv{5pt}{5}}
\makeatother

\begin{document}

\preprint{APS/123-QED}

\title{Generation of quantum phases of matter and finding a maximum-weight independent set of unit-disk graphs using Rydberg atoms} 

%----%----%----%----%----%----%----%----%----%----%----%----%----%----
\author{Ahmed M. Farouk~\orcidlink{0000-0002-6230-1234}}
\email[]{ahmed.farouk@azhar.edu.eg}
\affiliation{Faculty of Mechanics and Mathematics, Novosibirsk State University, 630090 Novosibirsk, Russia}
\affiliation{Laboratory of Nonlinear Resonant Processes and Laser Diagnostics, Rzhanov Institute of Semiconductor Physics SB RAS, 630090 Novosibirsk, Russia}
\affiliation{Department of Mathematics, Faculty of Science, Al-Azhar University, 11884, Cairo, Egypt}
%----%----%----%----%----%----%----%----%----%----%----%----%----%----
\author{I.I. Beterov~\orcidlink{0000-0002-6596-6741}}
\email[]{beterov@isp.nsc.ru}
\affiliation{Faculty of Physics, Novosibirsk State University, 630090 Novosibirsk, Russia}
\affiliation{Laboratory of Nonlinear Resonant Processes and Laser Diagnostics, Rzhanov Institute of Semiconductor Physics SB RAS, 630090 Novosibirsk, Russia}
\affiliation{Department of Laser Physics, Institute of Laser Physics SB RAS, 630090 Novosibirsk, Russia}
\affiliation{Faculty of Physical Engineering, Novosibirsk State Technical University, 630073 Novosibirsk, Russia}
%----%----%----%----%----%----%----%----%----%----%----%----%----%----
\author{Peng Xu~\orcidlink{0000-0001-8477-1643}}
\affiliation{State Key Laboratory of Magnetic Resonance and Atomic and Molecular Physics, Innovation Academy for Precision Measurement Science and Technology, Chinese Academy of Sciences, Wuhan 430071, China}
\affiliation{Department of Quantum Computing, Wuhan Institute of Quantum Technology, Wuhan 430206, China}
%----%----%----%----%----%----%----%----%----%----%----%----%----%----
\author{I.I. Ryabtsev~\orcidlink{0000-0002-5410-2155}}
\affiliation{Faculty of Physics, Novosibirsk State University, 630090 Novosibirsk, Russia}
\affiliation{Laboratory of Nonlinear Resonant Processes and Laser Diagnostics, Rzhanov Institute of Semiconductor Physics SB RAS, 630090 Novosibirsk, Russia}

%----%----%----%----%----%----%----%----%----%----%----%----%----%----
\date{\today}% It is always \today, today,
\begin{abstract}
Recent progress in quantum computing and quantum simulation of many-body systems with arrays of  neutral atoms using Rydberg excitation has provided unforeseen opportunities towards computational advantage in solving various optimization problems. The problem of a maximum-weight independent set of unit-disk graphs  is an example of an NP-hard optimization problem. It involves finding the largest set of vertices with the maximum sum of their weights for a graph which has edges connecting all pairs of vertices within a unit distance. This problem can be solved using quantum annealing with an array of interacting Rydberg atoms. For a particular graph, a spatial arrangement of atoms represents vertices of the graph, while the detuning from resonance at Rydberg excitation defines the weights of these vertices. The edges of the graph can be drawn according to the unit disk criterion.  Maximum-weight independent sets can be obtained by applying a variational quantum adiabatic algorithm. We consider driving the quantum system of interacting atoms to the many-body ground state using a non-linear quasi-adiabatic profile for sweeping the Rydberg detuning. We also propose using a quantum wire which is a set of auxiliary atoms of a different chemical element to mediate strong coupling between the remote vertices of the graph. We investigate this effect for different lengths of the quantum wire. We also investigate the quantum phases of matter realizing commensurate and incommensurate phases in one- and two-dimensional spatial arrangements of the atomic array.
\end{abstract}

%Use showkeys class option if keyword
\maketitle

%----%----%----%----%----%----%----%----%----%----%----%----%----%----%----%----%----%----%----%----%----%----%----%----%----%----%----%----%----%----%----%----%----%----%----%----%----%----%----%----%----%----%----%----%----%----%----%----%----%----%----%----%----%----%----%----%----%----%----%----%----%----%----%----%----%----%----%----%----%----%----%----%----%----%----%----%----%----%----%----%----%----%----%----%----%----%----%----%----%----%----%----%----%----%----%----%----%----%----%----%----%----%----%----%----%----%----%----%----%----%----%----%----%----%----%----%----%----%----%----%----%----%----%----%----%----%----%----%----%----%----%----%----%----%----%----%----%----%----%----%----%----%----%----%----%----%----%----%----%----%----%----%----%----%----%----%----%----%----%----%----%----%----%----%----%----%----%----%----%----%----%----%----%----%----%----%----%----%----%----%----%----%----%----%----%----%----%----%----%----%----%----%----%----%----%----%----%----%----%----%----%----%----%----%----%----%----%----%----%----%----%----%----%----%----%----%----%----%----%----%----%----%----%----%----%----%----%----%----%----%----
%%%%%%%

\section{Introduction}
In the realm of computer science~\cite{arora2009computational} there is a class of non-deterministic polynomial-time NP-hard problems that are notoriously challenging to solve efficiently. The algorithms designed for classical computers struggle to find solutions of such problems within a reasonable time frame/cost. With the development of the capabilities of quantum computing technologies there is a growing interest in finding useful applications of near-term quantum machines~\cite{wagner2023benchmarking}. Recent achievements in this field have allowed further exploration of the possibility of finding solutions of a number of NP-hard problems~\cite{kim2022rydberg}. In particular, recently it has been proposed to implement a quantum annealer with heteronuclear ultracold Rydberg atoms in optical lattices, featuring a highly controllable environment to explore the many-body adiabatic passage~\cite{glaetzle2017coherent}. 

\begin{figure*}[!]\centering
	\includegraphics[width=\textwidth]{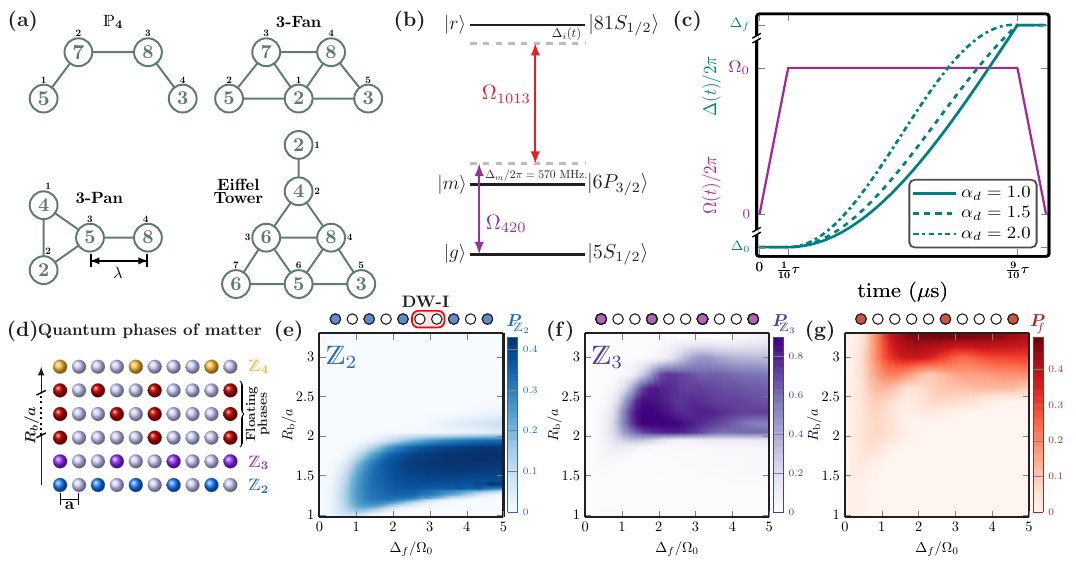}
	\caption{(a) Arbitrary undirected planar graphs being represented by Rydberg atoms. The positions of the vertices are defined as a function of graph spacing constant~$\lambda$ in Table~\ref{Table:AtomsPosition}. The  labels inside the circles represent weights of the corresponding vertices. The numbers outside the circles mark the position of each vertices. (b) Scheme of the atomic energy levels and laser excitation scheme  for $^{87}$Rb atoms. (c) Time profiles of two-photon Rabi frequency $\Omega(t)$ and  detuning from the Rydberg state $\Delta(t)$ for three different laser excitation courses with parameters $\alpha$ in Eq.~(\ref{Eq:DetuningFunction}): $\alpha_{d}=1.0$ (teal solid curve), $\alpha_{d}=1.7$ (teal dashed curve), and $\alpha_{d}=2.0$ (teal dash-dotted curve). Here $\Omega_0/2\pi=1.75$~\si{MHz}, $\Delta_{0}/2\pi=-6.0$~\si{MHz}, $\Delta_{f}/2\pi=+6.0$~\si{MHz}, and $\tau=5.0$~\si{\mu s}. (d) Scheme of quantum phases of matter represented by Rydberg excitations in the atomic array (shown as colored circles) with commensurate order $\mathds{Z}_2$, $\mathds{Z}_3$, and $\mathds{Z}_4$ and possible incommensurate floating phases between $\mathds{Z}_3$ and $\mathds{Z}_4$. Numerically calculated probability distributions (e) $\mathbf{P}_{\mathds{Z}_2}$, (f) $\mathbf{P}_{\mathds{Z}_3}$ and (g) $\mathbf{P}_{f}$ are plotted for ordered states $\mathds{Z}_2$ and $\mathds{Z}_3$ and an incommensurate floating phase, respectively, for an array of 10 atoms arranged in 1D with lattice spacing constant $a$~(\si{\mu m}). The red rectangle on top of graph (e) marks the position of a domain wall.}
	\label{Fig:SysDynamics}
\end{figure*}

In graph theory~\cite{bondy1982graph} we define a graph $G(V,E)$ with $V$ the set and $E$ the edges. A graph is called \textit{planar} if its vertices are embedded in the plane and its edges are not intersecting. Non-planar graphs represent a class of graphs where the edges between some of the vertices are intersecting and the vertices can be represented in three-dimensional space. A subset $S$ of a set $V$ is called an independent set if there are no couples of vertices of $S$ which are connected by the edge $E$. If there is no larger independent set $S'$ with $|S'|>|S|$, then $S$ is a maximum independent set (MIS). In other words, the size of the maximum independent set is the cardinality (number of vertices) of the largest independent set of the graph. Assigning a weight $w_v>0$ to each vertex $v\in V$ generalizes the MIS problem to a maximum-weight independent set (MWIS) problem, which can be solved by finding the independent set with the total weight $W=\sum_{v\in V}w_v$. Finding the solution for the MIS or MWIS is an NP-hard problem~\cite{arora2009computational}, meaning there is no known efficient algorithm to solve it for all possible graphs. However, there are approximate  classical algorithms and heuristics that can provide good solutions in practice. Examples of weighted graphs considered in our work are presented in Fig.~\ref{Fig:SysDynamics}(a).

%%%%%%%%%%%%%%%%%%%%%%%%%%%%%%%%%%%%%%%%%%%%%%%%%%%%%%%%%%%%%%%%%%%%%%%%%%%%%%%%

%%%%%%%%%%%%%%%%%%%%%%%%%%%%%%%%%%%%%%%%%%%%%%%%%%%%%%%%%%%%%%%%%%%%%%%%%%%%%%%%

Different approaches related to quantum computing can be used to solve both MIS and MWIS problems. Basically, there are two main approaches to driving the quantum system to the many-body ground state, which reveals a solution for MIS, MWIS, or other optimization problems. These approaches can be implemented on a physical platform of ultracold neutral atoms using advantages provided by specific properties of laser-driven atomic systems with Rydberg interactions~\cite{graham2022multi}.  The first one is a quantum approximate optimization algorithm (QAOA)~\cite{zhou2020quantum, munoz2023low, paradezhenko2023probabilistic} and its hybrid quantum iterative version~\cite{brady2023iterative, finvzgar2024quantum}. QuEra's quantum hardware processor, Aquila, has been used to find a solution for the max-cut problem and the IEEE 9-bus power grid graph state~\cite{bauer2024solving}. To physically implement the QAOA with neutral atoms, it is possible to use Rydberg excitation of the atomic ensemble by resonant laser pulses with varied  pulse phases and durations. These variational parameters are used to generate the desired many-body evolution. The other approach is a variational quantum adiabatic algorithm (VQAA), which was theoretically proposed in~\cite{farhi2000quantum}, optimized for adiabatic paths in~\cite{PRXQuantum.3.020347}, and physically implemented with atomic ensemble in~Refs.~\cite{scholl2021quantum, ebadi2021quantum, taylor2022simulation}. In this approach the detuning of a laser pulse exciting the atom to the Rydberg state is swept from the initial negative value to a large final positive value, while the Rabi frequency is kept constant. The VQAA has shown better performance compared with the QAOA due to the difficulty of finding the optimal parameters for implementing the QAOA with high depth circuits~\cite{ebadi2022quantum}. A comparison of the performance of the state-of-the-art classical solvers with the QAOA and VQAA for different optimization problems has been reported~\cite{lykov2023sampling, andrist2023hardness}. Recently, many researchers have proposed hybrid methods and algorithms for using neutral-atom architectures to find solutions of graph problems~\cite{coelho2022efficient,lu2024digital}. A protocol for solving hard combinatorial graph problems by combining  variational analog quantum computing and machine learning is assessed by quality score in Ref.~\cite{coelho2022efficient}. Moreover, in~Ref.\cite{lu2024digital} a hybrid digital-analog algorithm on Rydberg atoms shows the feasibility of the VQAA to the near-term implementation of quantum learning with the scalable architecture of arrays of neutral atoms.

%%%%%%%
Finding solutions for optimization problems as MIS/MWIS of unit-disk graphs is beneficial for practical applications such as network design, scheduling tasks~\cite{kose2017scheduling, finvzgar2023designing}, gene selection, and prediction of protein-protein interaction in bioinformatics~\cite{przulj2005graph}. Realizing the solutions of the MIS has been experimentally achieved using an atomic architecture with quantum wires for planar and non-planar graphs~\cite{kim2022rydberg}, transforming Platonic 3D planar graphs to 2D planar graphs~\cite{byun2022finding}, using 3D spatial arrangements of atoms to embed non unit-disk, non-planar or other more complex classes of graphs \cite{dalyac2023exploring}, and also for a large-size graph on the King’s lattice~\cite{kim2023quantum}. Construction of rigorous and challenging solutions for the MIS problem via the adiabatic approach was presented in Ref.~\cite{schiffer2023circumventing}. The framework for solving combinatorial optimization problems for non-unit-disk graphs was provided for  MIS and max-cut problems in Ref.~\cite{goswami2023solving} and adjusting local detunings on atoms to approximate the MIS was studied in Ref.~\cite{yeo2024approximating}. A generalized framework for solving the MWIS, and other optimization problems was discussed in Refs.~\cite{nguyen2023quantum, lanthaler2023rydberg}.
%----%----%----%----%----%----%----%----%----%----%----%----%----%----%----%----%----%----%----%----%----

In this paper, we study theoretically the problem of finding MIS/MWIS on unit-disk graphs using an array of ultracold neutral atoms and the VQAA. We propose a nonlinear quasiadiabatic profile of sweeping the detuning from resonance when exciting atoms into Rydberg states. We study the quantum phases of matter by analyzing commensurate $\mathds{Z}_n$-ordered states and incommensurate phases of matter in 1D and 2D spatial arrangements of atoms and we discuss the ability to construct a desired phase by controlling the system parameters. We study the effect of the sweeping rate on the generation of $\mathds{Z}_{n}$ ordered states and the detection of domain walls and we calculated the critical detuning which is related to the minimum weight of each vertex in weighted graphs.

 We investigate the solutions of the MIS and MWIS for planar and nonplanar graphs. Then we consider a dual-species quantum architecture for mediating interaction between far distant atoms by a quantum wire composed of atoms from different chemical elements. Quantum wires also can be used to consider non-unit-disk graphs. Dual-species quantum architectures are advantageous to avoid crosstalk between the graph atoms and wire atoms during the measurement process, following the initial proposal~\cite{PhysRevA.92.042710}. This architecture also provides opportunities for blockade enhancement and additional flexibility of interaction energies by dual-species F\"{o}rster resonances. Moreover, we investigate the cost of employing quantum wires of different lengths on mediating the interaction between distant atoms for finding the MWIS.

The paper is organized as follows: In Sec.~\ref{Sec:QSystem} we discuss the Hamiltonian of the atomic system and the procedure used for annealing the system to the desired many-body ground state.  We show the dynamics of the states of a laser-driven single atom and of a spatially arranged ensemble of atoms representing a graph. In Sec.~\ref{Sec: QPT} we study the quantum phase transition of commensurate and incommensurate phases, and discuss the ability to realize a floating incommensurate phase.  In Sec.~\ref{Sec:PlanarGraphs} we discuss results of MISs and MWIS for our simulation for planar graphs. In section~\ref{Sec:NonPlanarGraph}, we find the MISs and MWIS of a nonplanar graph representing a Johnson solid $J_{14}$, whose facets are different regular polygons in a three-dimensional array. The use of a heteronuclear quantum wire to mediate the interaction between distant atoms is discussed in Sec.~\ref{Sec:QuantumWire}. We summarize and discuss our results in Sec.~\ref{Sec:Conclusion}.

%----%----%----%----%----%----%----%----%----%----%----%----%----%----%----%----%----%----%----%----%----%----%----%----%----%----%----%----%----%----%----%----%----%----%----%----%----%----%----%----%----%----%----%----%----%----%----%----%----%----%----%----%----%----%----%----%----%----%----%----%----%----%----%----%----%----%----%----%----%----%----%----%----%----%----%----%----%----%----%----%----%----%----%----%----%----%----%----%----%----%----%----%----%----%----%----%----%----%----%----%----%----%----%----%----%----%----%----%----%----%----%----%----%----%----%----%----%----%----%----%----%----%----%----%----%----%----%----%----%----%----%----%----%----%----%----%----%----%----%----%----%----%----%----%----%----%----%----%----%----%----%----%----%----%----%----%----%----%----%----%----%----%----%----%----%----%----%----
\section{Quantum system\label{Sec:QSystem}}

%%%%%%%%----%----%----%----%----%----%----%----%----%----%----%----%----%----%----%----%----%----%----%----%----%----%----%----%----%----%----%----%----%----%----%----%----%----%----%----%----%----%----%----%----%----%----%----%----%----%----%----%----%----%----%----%----%----%----%----%----%----%----%----%----%----%----%----%----%----%----%----%----%----%----%----%----%----%----%----%----%----%----%----%----%----%----%----%----%----%----%----%----%----%----%----%----%----%----%----%----%----%----%----%----%----%----%----%----%----%----%----%----%----%----%----%----%----%----%----%----%----%----%----%----%----%----%----%----%----%----%----%----%----%----%----%----%----%----%----%----%----%----%----%----%----%----%----%----%----%----%----%----%----%----%----%----%----%----%----%----%----%----%----%----%----%----%----%----%----%----%----

The spatially arranged array of atoms excited into the Rydberg states can be represented by a mathematical graph $G(V,E)$ in which the vertices $V$ represent the atoms, and the edges $E$ represent the pairwise interaction between atoms. The governing Hamiltonian of this system can be written as
\begin{equation}
 \mathcal{\hat{H}}=\mathcal{\hat{H}}_{q}+\mathcal{\hat{H}}_{c},
\end{equation}
where $\mathcal{\hat{H}}_{q}$ is the quantum driver Hamiltonian which is composed of off-diagonal operators, and $\mathcal{\hat{H}}_{c}$ is the cost Hamiltonian~\cite{PhysRevA.93.062312, kivlichan2017bounding}. Minimizing the cost parameters is an ultimate goal. The scheme of atomic energy levels is shown in Fig.~\ref{Fig:SysDynamics}(b). These two basic ingredients of the system Hamiltonian are given by
\begin{equation}
\mathcal{\hat{H}}_{q}=\frac{1}{2} \sum_{i}^{V} \bigr[\Omega(t) | g \rangle_{i} \langle r | + \text{H.c.} \bigr],
\end{equation}
\begin{equation}
\mathcal{\hat{H}}_{c}=-\sum_{i}^{V}  \Delta_{i}(t)~\hat{n}_{i}+\sum_{i<j}^{E} \frac{\mathcal{C}_{6}}{|R_i-R_j|^{6}} \hat{n}_{i} \hat{n}_{j},
\end{equation}
where $|g\rangle=|5S_{1/2},F=2\rangle$ is the ground state of the trapped  $^{87}$Rb atom, $|r\rangle=|81S_{1/2}, m_j=1/2\rangle$ is the Rydberg state, and the operator $\hat{n}_i=|r\rangle_{i}\langle r|$ is the projector to the Rydberg state of $i$th atom. In addition, $\Omega(t)$ is a time-dependent effective two-photon Rabi frequency for the $|g\rangle \to |r\rangle$ transition with the maximum  value $\Omega_0=\Omega_{420}~\Omega_{1013}/2\Delta_{m}$, where $\Omega_{420}/2\pi=40$~\si{MHz} and $\Omega_{1013}/2\pi=50$~\si{MHz} are the one-photon Rabi frequencies of laser radiation which couples $|g\rangle$ to $|r\rangle$ through the intermediate state $|m\rangle=|6P_{3/2},F=2,m_F=2\rangle$, and $\Delta_{m}/2\pi=570$~\si{MHz} is the detuning from the intermediate state. The values of Rabi frequencies are chosen to excite a single atom to the Rydberg state according to the considered time scale with high fidelity. Further, $\Delta_{i}(t)$ is the detuning from the two-photon resonance with the Rydberg state $|r\rangle$ of the atom or vertex $i$. For the MIS problem we set identical detunings for all atoms $\Delta_i(t)\equiv \Delta(t)$. Later, for the MWIS problem we will select detuning for each atom individually. We used the following shapes of time profiles $\Omega(t)$ and $\Delta(t)$ [see Fig.~\ref{Fig:SysDynamics}(c)]:
\begin{equation}
	\Omega(t) = \left\{
	\begin{array}{ll}
		\Omega_0 (\frac{t}{t_{f_1}}), & 0\leq t \leq t_{f_1}\\
		\Omega_0, & t_{f_1} < t \leq t_{f_2}\\
		 \Omega_0\frac{(t-\tau)}{(t_{f_2}-\tau)}, & t_{f_2}< t \leq \tau,
	\end{array}
	\right.
\end{equation}

\begin{equation}\label{Eq:DetuningFunction}
	\Delta(t) = \left\{
	\begin{array}{ll}
		\Delta_{0}, & 0 \leq t < t_{f_1}\\
		\mathds{A} \sin^{2}\left(\alpha_{d}\dfrac{t-t_{f_{1}}}{\tau}\right)+\Delta_{0}, & t_{f_1} \leq t < t_{f_2}\\
		\Delta_{f}, & t_{f_2} \leq t \leq \tau.\\
	\end{array}
	\right.
\end{equation}
Here $\tau = 5$~\si{\mu s} is a quantum annealing time, $t_{f_1}=\frac{1}{10}\tau$, and $t_{f_2}=\frac{9}{10}\tau$. In addition, $\Delta_{0}$ and $\Delta_{f}$ are the initial and final values of Rydberg detuning. The amplitude $\mathds{A}=(\Delta_{f} - \Delta_{0})\csc^{2} (\alpha_{d} \frac{t_{f_{2}} - t_{f_{1}}}{\tau})$  guarantees that $\Delta$ will not exceed the predefined maximum of $\Delta_{f}$ for any possible value of the parameter $\alpha_{d}>0$, which controls the course of detuning [see Fig.~\ref{Fig:SysDynamics}(c)].  The nonlinear quasi-adiabatic time profile of the detuning at a constant value of Rabi frequency $\Omega_0$ minimizes non-adiabatic excitations~\cite{PhysRevA.94.062307}; almost the same profile was used in~\cite{PRXQuantum.5.020362} as the parameter of adiabatic path, which minimizes the cost Hamiltonian. To find the MISs of graphs with unweighted vertices,  the value of  $\Delta_{f}$ is kept constant. However, for graphs with weighted vertices the maximum value of Rydberg state detuning $\Delta_{f}$ will define each vertex weight of the corresponding atom and should be selected individually for each atom, as we discuss in Sec.~\ref{Sec:MWIS-PlanarGraphs}.

In the regime of the Rydberg blockade~\cite{browaeys2020many}, when simultaneous laser excitation of two Rydberg atoms located at small interatomic distances becomes impossible, Rydberg interactions within an atomic array result in complex phases and phase transitions. For  1D arrays of interacting atoms the phase of a quantum system can be structured into $\mathds{Z}_{\text{n}}$-ordered states ($n\geq2$ is the number of  sites separating neighboring Rydberg atoms), which are a class of commensurate phases.  Spatial arrangements of Rydberg excitations in different 1D atomic arrays for the quantum phase of matter with commensurate order $\mathds{Z}_2$, $\mathds{Z}_3$, and $\mathds{Z}_4$ and other possible incommensurate floating phases between $\mathds{Z}_3$ and $\mathds{Z}_4$ are shown in Fig.~\ref{Fig:SysDynamics}(d). 
The  commensurate phases $\mathds{Z}_2$ and $\mathds{Z}_3$, numerically calculated for a 1D array of $10\times 1$ atoms, are shown in Figs.~\ref{Fig:SysDynamics}(e), \ref{Fig:SysDynamics}(f), and \ref{Fig:SysDynamics}(g), respectively. The details of the numeric calculations will be given below.

\begin{figure}[htb!]\centering
	\includegraphics[width=\columnwidth]{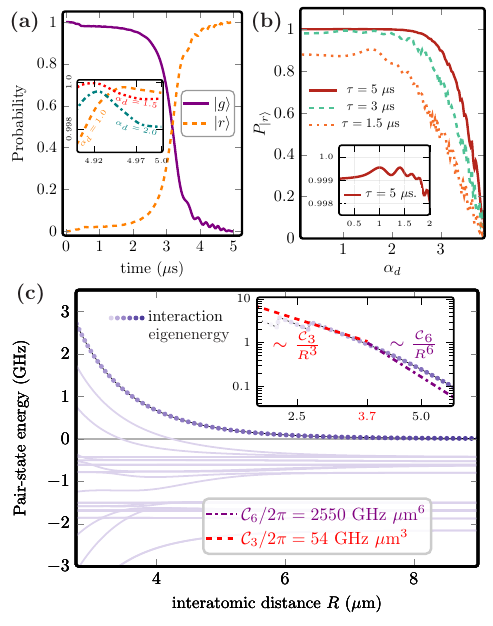}
	\caption{(a) Time dynamics of the probabilities of  ground and Rydberg states during Rydberg excitation of a single atom for different values of $\alpha_d$. (b) Probability of exciting a single atom to the Rydberg state $P_{|r\rangle}$ as a function of the detuning course parameter $\alpha_{d}$ for different values of annealing time $\tau$. (c) Pair-state interaction energy for two $^{87}$Rb, $|81S_{1/2}, m_j=1/2\rangle$ Rydberg atoms. Teal solid curves are the eigenenergies of the interaction matrix after diagonalisation, calculated using the alkali Rydberg calculator~\cite{vsibalic2017arc}, with $\theta=0$, $\phi=0$, $\delta n=2$, and $\delta l=1$. Curves are the fitting approximations for short-range dipole-dipole (red dashed curve) and long-range van der Waals (violet dash-dotted curve) regimes, showing that $\mathcal{C}_3/2\pi=54$~\si{GHz}~\si{\mu m^3}, $\mathcal{C}_6/2\pi=2550$~\si{GHz}~\si{\mu m^6}, $R_{\text{\tiny LR}}=2.0$~\si{\mu m}, and $R_{\text{\tiny vdW}}=3.7$~\si{\mu m}. The inset shows a log-log plot of the fitting values and the corresponding eigenenergy. }
	\label{Fig:LevelScheme}
\end{figure}

The numerically calculated probabilities to excite a single atom to the Rydberg state using the quasiadiabatic profile from Eq.~(\ref{Eq:DetuningFunction}) is shown in Fig.~\ref{Fig:LevelScheme}(a) with a probability of $0.9995$ for the Rydberg state $|r\rangle$. Here, for simplicity, we neglect spontaneous decay of the excited states, and the fidelity is limited purely by the nonadiabatic dynamics of the quantum system. The inset in Fig.~\ref{Fig:LevelScheme}(a) shows the effect of different courses $\alpha_{d}$ of sweeping the detuning on the calculated probability of the single-atom  Rydberg excitation. In Fig.~\ref{Fig:LevelScheme}(b), we plot the probability of exciting single atom to Rydberg state as a function of the detuning course parameter $\alpha_{d}$ for different annealing times. We chose $\alpha_d=1$ in our calculations, since it provides maximum excitation probability as shown in the insets of Figs.~\ref{Fig:LevelScheme}(a) and \ref{Fig:LevelScheme}(b).

%----%----%----%----%----%----%----%----%----%----%----%----%----%----%----%----%----%----%----%----%----%----%----%----%----%----%----%----%----%----%----%----

%----%----%----%----%----%----%----%----%----%----%----%----%----%----%----%----%----%----%----%----%----%----%----%----%----%----%----%----%----%----%----%----
%----%----%----%----%----%----%----%----%----%----%----%----%----%----%----%----

For any two vertices $i, j$ ($i \neq j$) on a graph, which are represented by two atoms separated by an edge $E$, defined by interatomic distance $R=|R_{i}-R_{j}|$, the interaction between atoms can be either a short-range dipole-dipole interaction of approximately $\mathcal{C}_{3}/R^{3}$  for $R<R_{\text{\scalebox{0.8}{vdW}}}$ or a long-range van der Waals (vdW)  of approximately $\mathcal{C}_{6}/R^{6}$ interaction for $R_{\text{\scalebox{0.8}{vdW}}} < R$. The interatomic distance should be much larger than the Leroy radius $R_{\text{\scalebox{0.8}{LR}}}$, which marks the minimum interatomic distance between two atoms to satisfy Leroy-Bernstien theory~\cite{leroy1970dissociation}.  The vdW radius  $R_{\text{\scalebox{0.8}{vdW}}}$ characterizes the border between different interaction regimes. It depends on the structure and properties of the atomic energy levels of a particular chemical element and can be significantly increased in cases of an asymmetric homonuclear or heteronuclear Rydberg interaction in the vicinity of F\"{o}rster resonances~\cite{Farouk2023parallel, PhysRevA.92.042710}. In Fig.~\ref{Fig:LevelScheme}(c) we show the eigenenergies of the interaction Hamiltonian for two atoms excited symmetrically to the Rydberg state $|81S_{1/2}, m_j=1/2 \rangle$. We used  ARC~\cite{vsibalic2017arc} for calculations. The dominant regime for interatomic distances $R>R_{\text{\tiny vdW}}=3.7$~\si{\mu m} is the vdW interaction with $\mathcal{C}_6/2\pi=2550$~\si{GHz}~\si{\mu m^6}. The Rydberg blockade occurs when $R<R_{\text{\tiny b}}=\left(\mathcal{C}_6/\Omega_{0}\right)^{1/6}$, where $R_{\text{\tiny b}}$ is a blockade radius.

We use a time-dependent Schr\"{o}dinger equation $i \hbar\frac{\partial}{\partial t} |\psi\rangle=\mathcal{\hat{H}}|\psi\rangle$ to calculate the time dependences of probabilities in Fig.~\ref{Fig:SysDynamics}(c) for any graph with $N \geq 8$ vertices and to plot the phase diagrams of $\mathds{P}_4$ and 7-Pan graphs in Figs.~\ref{Fig:SysDynamics}(a), and \ref{Fig:QWire}(d), respectively. We perform a Monte Carlo simulation with the Lindblad master equation considering the radial positional fluctuations $\delta_{\text{\tiny R}}=\sqrt{\delta_{\text{\tiny x}}^2+\delta_{\text{\tiny y}}^2}$ of each atom arising from fluctuations in trapping power. Also, the axial positional fluctuations $\delta_{\text{\tiny z}}$ arising from the nonzero temperature of trapped atoms are considered in the calculations. The Lindblad master equation is written as
\begin{equation}\label{Eq:LindbladMaster}
\frac{d}{dt}\hat{\rho}=-\frac{i}{\hbar} \left[ \mathcal{\hat{H}},\hat{\rho} \right] + \sum_{i}^{V}\left( \hat{L}_{i} \hat{\rho} \hat{L}_{i}^{\dagger} - \frac{1}{2} \left\{ \hat{L}_{i}^{\dagger} \hat{L}_{i},\hat{\rho} \right\} \right),
\end{equation}
where $\hat{\rho}$ is the density matrix and $\hat{L}_{i} = \sqrt{\frac{1}{2} \gamma_{m} } \hat{n}_{i}$ is the jumping operator describing the dissipative processes in the system, with $\gamma_{m}=1/\tau_{|m\rangle}$ the decay of the intermediate state $|m\rangle$ with lifetime $\tau_{|m\rangle}=0.118$~\si{\mu s}. 

\section{Quantum phase transitions \label{Sec: QPT}}

The solution for the MIS and MWIS for unit-disk graphs can be found in the ordered phases of $\mathds{Z}_2$ or $\mathds{Z}_3$, depending on the spatial arrangement of the graph vertices. In this section, we study the commensurate (or ordered) and incommensurate states for a linear arrangement of atoms,  with the of obtaining a clear understanding of the regimes and stability of solutions.
%----%----%----%----%----%----%----%----%----%----%----%----%----%----%----%----%----%----%----%----%----%----%----%----%----%----%----%----%----%----%----%----
%----%----%----%----%----%----%----%----%----%----%----%----%----%----%----%----%----%----%----%----%----
\begin{figure}[t]\centering
	\includegraphics[width=\columnwidth]{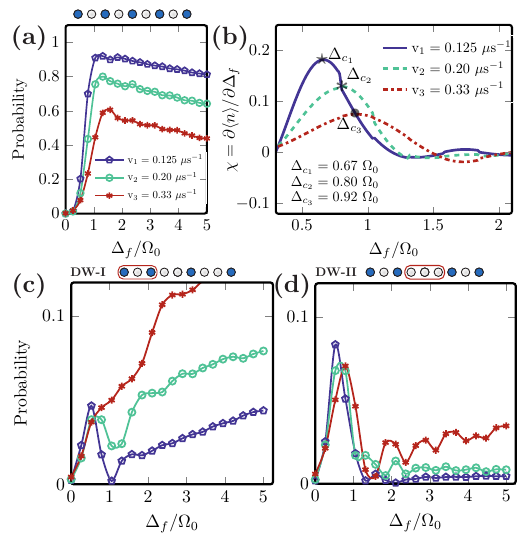}
	\caption{(a) Probability of obtaining the ordered state $\mathds{Z}_2$ for an array of $9 \times 1$ atoms as a function of the dimensionless parameter $\Delta_{f}/\Omega_{0}$ for different values of the annealing time $\tau$ by considering $R_{\text{\tiny b}}/a=1.5$~\si{\mu m}. (b) Susceptibility $\chi=\partial \langle n \rangle / \partial \Delta_{f}$ for different sweeping rates, showing the critical detuning $\Delta_{c}$ as the value of detuning corresponds to the maximum value of $\chi$. Also shown as the probability of obtaining (c) the first, and (d) the second type of domain wall, in a $\mathds{Z}_n$ chain of atoms.}
	\label{Fig:DomainWall-2}
\end{figure}
%----%----%----%----%----%----%----%----%----%----%----%----%----%----%----%----%----%----%----%----%----%----%----%----%----%----%----%----%----%----%----%----
%----%----%----%----%----%----%----%----%----%----%----%----%----%----%----%----

In the atomic arrays, a quantum phase transition into the $\mathds{Z}_{n}$-ordered state was realized experimentally in the 1D array~\cite{bernien2017probing, keesling2019quantum}, on a 1D ring~\cite{fang2024probing}, and on a 2D chequerboard phase~\cite{ebadi2021quantum}, which enabled the investigation of the quantum Kibble-Zurek mechanism (QKZM)~\cite{kibble1976topology, zurek1985cosmological} and the critical dynamics of ordered states. The QKZM provides a solid understanding of the nonequilibrium dynamics of cosmological, particle, and condensed-matter systems~\cite{PhysRevLett.116.155301}. Ordered states are beneficial for creating exotic states of matter with topological order, such as a quantum spin liquid~\cite{semeghini2021probing}.  The transition from the disordered phase to the ordered phase takes place at a specific value of the Rydberg detuning $\Delta_{c}$ depending on the phase of the transition and the value of the sweeping rate. Here $\Delta_{c}$ is called the critical detuning. The QKZM of commensurate phases of $\mathds{Z}_{\text{n}}$-ordered states of an atomic array was studied earlier in Refs.~\cite{PhysRevLett.95.105701, bernien2017probing, keesling2019quantum, ebadi2021quantum, fang2024probing} for equal  maximum values of Rydberg detuning  for all atoms in the array. Consequently, the value of critical detuning $\Delta_{c}$ is related to the number of separated sites~$n$ of $\mathds{Z}_{\text{n}}$-ordered states. Therefore, the parameters of critical dynamics, such as critical length and critical scaling exponents, which characterize the Ising universality class and the QKZM, can be obtained. Also, critical incommensurate phases (floating phases) were theoretically predicted in a 1D Rydberg array~\cite{PhysRevB.29.239, rader2019floating, PhysRevResearch.4.043102} and experimentally observed in 2D ladder Rydberg arrays~\cite{zhang2024probing}. 

To keep the interaction between atoms in the vdW regime, we restrict our calculations to $1\leq R_{b}/a\leq3.2$. For this restriction we find that the realization of $\mathds{Z}_4$ is not conceivable. In our numeric simulations of the commensurate phases $\mathds{Z}_2$ and $\mathds{Z}_3$ of the 1D array of $10\times 1$ atoms, which are shown in Figs.~\ref{Fig:SysDynamics}(e)-\ref{Fig:SysDynamics}(g), we have four and three spins for the $\mathds{Z}_2$ and $\mathds{Z}_3$, respectively. For $\mathds{Z}_2$-ordered state, the interruption of the ordering sequence is observed [which is indicated by the red rectangle in Fig.~\ref{Fig:SysDynamics}(e)]. This interruption of the ordering is called a \textit{domain wall} which occurs in different positions of the array. Domain walls are identified as having either one atom at the edge of the array in ground state or two neighboring atoms in the same state~\cite{bernien2017probing}. 

Fig.~\ref{Fig:DomainWall-2} illustrates the effect of the sweeping rate $\text{v}$ on the probability of finding the ordered state $\mathds{Z}_{2}$. In an ideal case without domain walls, as shown in Fig.~\ref{Fig:DomainWall-2}(a), it is clear that with increasing sweeping rate the maximum possible probability of the ordered state decreases. In Fig.~\ref{Fig:DomainWall-2}(b) we plot the susceptibility $\chi=\partial\langle n\rangle/\partial\Delta_{f}$, which is calculated by interpolating the numerical data and then differentiation. The maximum value of the susceptibility corresponds to the critical value of detuning $\Delta_{c}$. From Fig.~\ref{Fig:DomainWall-2}(b) it is clear that with an increase of the sweeping rate the value of $\Delta_{c}/\Omega_{0}$ shifts to higher values.   This is analogous to results of Ref.~\cite{fang2024probing}. A  pulse profile with moderate sweeping rate can minimize the detuning parameter in the cost Hamiltonian. To find a solution for the MWIS, in Sec.~\ref{Sec:MWIS-PlanarGraphs} we define the weight $w_i>\Delta_{c}$ of a vertex $i$. For quantum annealing time $\tau=5$~\si{\mu s} we select the weights following the condition $w_i>0.8~\Omega_{0}=2\pi\times1.4$~\si{MHz} and keep the lattice spacing constant at approximately $a\simeq \frac{1}{1.5} R_{\text{\tiny b}}$. 

%----%----%----%----%----%----%----%----%----%----%----%----%----%----%----%----%----%----%----%----%----%----%----%----%----%----%----%----%----%----%----%----%----%----%----%----%----%----%----%----
\begin{figure}[t]\centering
	\includegraphics[width=\columnwidth]{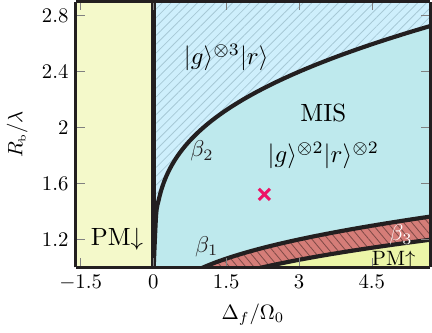}
	\caption{Phase diagram of the $\mathds{P}_4$ graph showing the regions of dominant phases. The antiferromagnetic phase where atoms are singly excited to the Rydberg state is bounded by $0< \Delta_{f}/\Omega_{0} \leq \beta_{2}(R_{\text{\tiny b}}/\lambda)$. The MIS phase is dominant in the region $\beta_{1}(R_{\text{\tiny b}}/\lambda) < \Delta_{f}/\Omega_{0} <\beta_{2}(R_{\text{\tiny b}}/\lambda)$. There are two PM phases: (i) the PM$\downarrow$ phase of ground states is bounded by $\Delta_{f}/\Omega_{0} < 0$, and (ii) the PM$\uparrow$ phase, which is bounded by $ \Delta_{f}/\Omega_{0} > \beta_{1}(R_{\text{\tiny b}}/\lambda)$. The phase bounded within $\beta_{1}(R_{\text{\tiny b}}/\lambda) < \Delta_{f}/\Omega_{0} < \beta_{3}(R_{\text{\tiny b}}/\lambda)$ is where three atoms can be excited to the Rydberg state simultaneously.}
	\label{Fig:PhaseDiagramP4}
\end{figure}

%----%----%----%----%----%----%----%----%----%----%----%----%----%----%----%----%----%----%----%----%----%----%----%----%----%----%----%----%----%----%----%----%----%----%----%----%----%----%----%----%----%----%----%----%----%----%----%----

The construction of the $\mathds{Z}_2$-ordered state with a defect (domain wall) induced by an ancilla and optimizing the driving fields on QuEra's quantum hardware Aquila was performed in Ref.~\cite{balewski2024engineering}.  The transition from the $\mathds{Z}_2$-ordered to the $\mathds{Z}_3$-ordered crystalline state occurs at $R_{b}\simeq 2 a$. In our simulation, we find that the incommensurate phases between $\mathds{Z}_2$ and $\mathds{Z}_3$, for a 1D spatial arrangement do not exist, as concluded in~\cite{keesling2019quantum}, despite the fact that they were predicted in Ref.~\cite{PhysRevB.69.075106}. According to our calculations, in the range of validity of the vdW interaction, an incommensurate floating phase emerged in the regime of $2.6\leq R_{b}/a\leq 3.2$ and $\Delta_{f}/\Omega_{0}\geq1$. This is supposed to be a floating phase between $\mathds{Z}_{3}$ and $\mathds{Z}_{4}$.

There are two types of domain walls depending on the number of interrupted sites~\cite{PhysRevA.98.023614}. Domain walls of type~I and type~II  include two and three interrupted sites, respectively. High sweep rates reduce the probability of obtaining an ideal case for the ordered state $\mathds{Z}_2$, as illustrated in Fig.~\ref{Fig:DomainWall-2}(a). Therefore, the possibility of obtaining a chain of atoms with interrupted sites, or domain walls, increases. In Figs.~\ref{Fig:DomainWall-2}(c) and \ref{Fig:DomainWall-2}(d) we show the effect of increased sweeping rates on the probability of obtaining an interrupted $\mathds{Z}_n$ state with type~\textrm{I} and type~\textrm{II} domain walls, respectively.

%----%----%----%----%----%----%----%----%----%----%----%----%----%----%----%----%----%----%----%----%----%----%----%----%----%----%----%----%----%----%----%----%----%----%----%----%----%----%----%----%----%----%----%----%----%----%----%----%----%----%----%----%----%----%----%----%----%----%----%----%----%----%----%----%----%----%----%----%----%----%----%----%----%----%----%----%----%----%----%----%----%----%----%----%----%----%----%----%----%----%----%----%----%----%----%----%----%----%----%----%----%----%----%----%----%----%----%----%----%----%----%----%----%----%----%----%----%----%----%----
%----%----%----%----%----%----%----%----%----%----%----%----%----%----%----%----%----%----%----%----%----%----%----%----%----%----%----%----%----%----%----%----%----%----%----%----%----%----%----%----%----%----%----%----%----%----%----%----%----%----%----%----%----%----%----%----%----%----%----%----%----%----%----%----%----%----%----%----%----%----%----%----%----%----%----%----%----%----%----%----

\section{Planar Graphs\label{Sec:PlanarGraphs}}
%In this section, we discuss the results of driving atomic system of an ensemble of graphs shown in Fig.~\ref{Fig:GraphsArb} from the ground state to the many-body ground state.

\subsection{Maximum Independent Set (MIS)}
%----%----%----%----%----%----%----%----%----%----%----%----%----%----%----%----%----%----%----%----%----%----%----%----%----%----%----%----%----%----%----%----%----%----%----%----%----%----%----%----

\begin{figure*}[!htb]\centering
	\includegraphics[width=\textwidth]{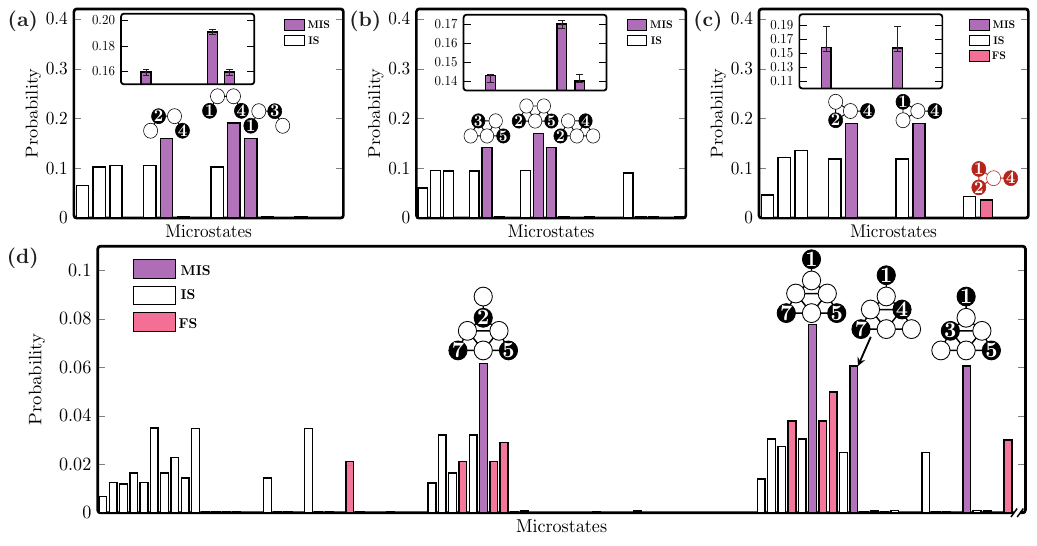}
	\caption{Probability distributions indicating the MIS solutions (violet bars) of the graphs (a) $\mathds{P}_{4}$, (b) three-fan, (c) three-pan, and (d) Eiffel-tower, for $\Delta_{0}/2\pi =-6.0$~\si{MHz}, $\Delta_{f}/2\pi=4.0$~\si{MHz}, $\tau=5.0$~\si{\mu s}, and the graph spacing constant $\lambda=6.0$~\si{\mu m}. The insets show the errors bars resulting from considering positional (radial and axial) fluctuations, $\delta_{\text{\tiny R}}=\pm 0.447$~\si{\mu m} ($\delta_{\text{\tiny x}}=\delta_{\text{\tiny y}}=0.1$~\si{\mu m}), and $\delta_{\text{\tiny z}}= \pm 0.6$~\si{\mu m}. White (red) bars correspond to the independent sets (frustrated sets). In (c), only the first 90 states of total $2^7=128$ states are shown. The probabilities of other states are infinitely small. A black-filled circle indicates a  weightless vertex whose corresponding atom is excited to the Rydberg state with labels indicating the vertex number. Open circles correspond to atoms in the ground state.}
	\label{Fig:MIS-Planar-Graphs}
\end{figure*} 
%----%----%----%----%----%----%----%----%----%----%----%----%----%----%----%----%----%----%----%----%----%----%----%----%----%----%----%----%----%----%----%----%----%----%----%----%----%----%----%----
The maximum independent sets of unweighted planar graphs are obtained  for the maximum Rydberg detuning $\Delta_{f}$ equal for all atoms resembling the graph, as shown in Fig.~\ref{Fig:SysDynamics}(a). Here we omit the weights of the vertices.  The graph spacing constant $\lambda$ defines the distance between vertex $i$ and the nearest vertex $j$ ($i \neq j$) and is an alternative to the lattice constant $a$, which is shown in Fig.~\ref{Fig:SysDynamics}(d).

The phases of the $\mathds{P}_4$  graph, shown in Fig.~\ref{Fig:SysDynamics}(a), could be different from the phases of an 1D atomic array due to differences in the energies of all pairwise interactions in a two-dimensional graph. In Fig.~\ref{Fig:PhaseDiagramP4} the phase diagram of the $\mathds{P}_4$ graph shows regions of dominant phases. The phase diagram is obtained by parametrizing the Hamiltonian via the ratios  $R_{b}/\lambda$ and  $\Delta_{f}/\Omega_{0}$. The positions of the graph vertices are given in Table~\ref{Table:AtomsPosition} as a function of the graph spacing constant~$\lambda$. Fluctuations of positions of atoms, laser noise, and spontaneous emissions are neglected. The boundaries between different phases can be fitted by functions $\beta_{i}(\frac{R_{b}}{\lambda})=\frac{1}{4} \mathcal{C}_{6}^{\text{\scalebox{0.7}{Rb-Rb}}} e^{-\xi_{i} R_{b}/\lambda}$\footnote{We consider here the numerical value of $\mathcal{C}_{6}$.}  where the parameters $\xi_{1}=1.4$, $\xi_{2}=2.8$, and $\xi_{3}=1.23$ define three different curves $\beta_{1}$, $\beta_{2}$, and $\beta_{3}$, shown in Fig.~\ref{Fig:PhaseDiagramP4}. There are two different regions where paramagnetic (PM) phases are dominant: (i) The PM$\downarrow$ phase is  located where $\Delta_{f}< 0$ and only the ground state of all four atoms $|g\rangle^{\otimes 4}$ can be found in this region of parameters and (ii) the PM$\uparrow$ phase is located where $\Delta_{f}/\Omega_{0} > \beta_{3}$, where all atoms are excited to the Rydberg state. The bounded phase within $\beta_{3} \leq \Delta_{f}/\Omega_{0} \leq \beta_{1}$ is a PM$\uparrow$-like phase where three atoms can be excited to the Rydberg state. For $0 < \Delta_{f}/\Omega_{0} \leq \beta_{2}$, we observe a phase with  only one atom  excited to the Rydberg state due to the strong interaction between atoms. The state of the system in this phase can be written as $|\psi\rangle=\sum_{l\neq i,j,k}^{N} |g\rangle^{\otimes3}_{i,j,k}|r\rangle_{l}$. The region, bounded between $\beta_{1}$ and $\beta_{2}$ is an antiferromagnetic phase revealing the MISs. 

In Fig.~\ref{Fig:MIS-Planar-Graphs} we show the calculated probability distributions of all possible states of an atomic system after being driven to the many-body ground state. We set the values of the graph spacing constant $\lambda=7.0$~\si{\mu m} and the maximum value of  Rydberg detuning $\Delta_{f}/\Omega_{0} = 2.28$, which are compromised values for finding MISs, as pointed out in Fig.~\ref{Fig:PhaseDiagramP4}. These values of $\lambda$ and $\Delta_{f}$ will be constant for all graphs in Fig.~\ref{Fig:MIS-Planar-Graphs}. We consider finite lifetimes of intermediate excited and Rydberg states using Lindblad equation, and perform a Monte Carlo simulation to take into account fluctuations of atomic positions.
%----%----%----%----%----%----%----%----%----%----%----%----%----%----%----%----%----%----%----%----%----%----%----%----%----%----%----%----%----%----%----%----%----%----%----%----%----%----%----%----%----%----%----%----%----%----%----%----%----%----%----%----%----%----%----%----

\begin{figure*}[!]\centering
	\includegraphics[width=\textwidth]{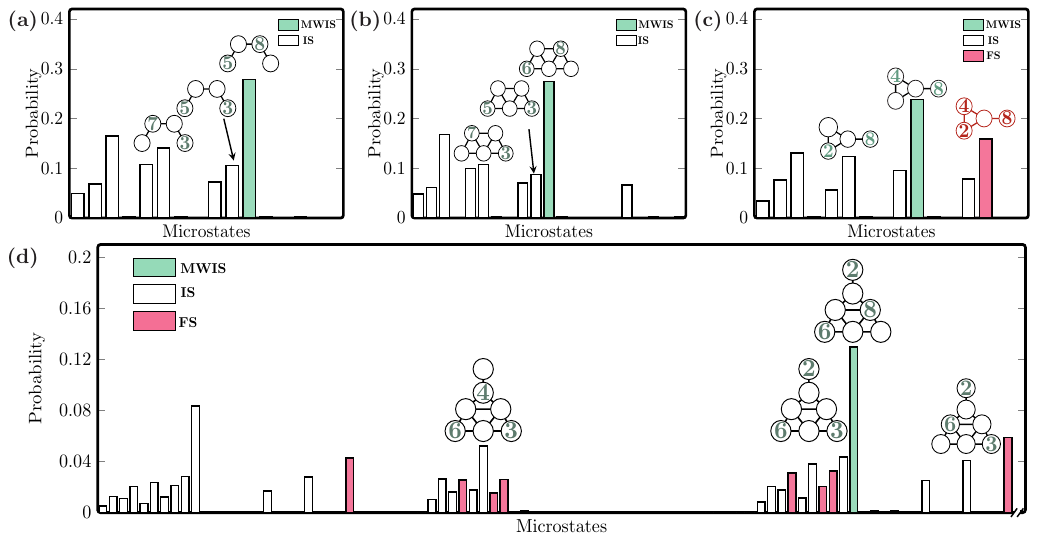}
	\caption{Probability distributions indicating the MWIS solutions (green bars) of the corresponding graphs in Fig.~\ref{Fig:SysDynamics}(a) with the same values of parameters, as in Fig.~\ref{Fig:MIS-Planar-Graphs}, but with different values of $\Delta_{f}$ [see the text for details]. White (red) bars correspond to the independent sets (frustrated sets). A circle with a number  indicates that the corresponding atom is excited to the Rydberg state, with the light green label indicating the weight of this vertex and the open circles representing atoms in the ground state.}
	\label{Fig:MWIS-Planar-Graphs}
\end{figure*} 

%----%----%----%----%----%----%----%----%----%----%----%----%----%----%----%----%----%----%----%----%----%----%----%----%----%----%----%----%----%----%----%----%----%----%----%----%----%----%----%----%----%----%----%----%----%----%----%----%----%----%----%----%----%----%----%----
The MISs of the $\mathds{P}_4$ graph are \textbf{\{}\{1, 3\}, \{1, 4\}, \{2, 4\}\textbf{\}}. These states are shown in Fig.~\ref{Fig:MIS-Planar-Graphs}(a) as positions of Rydberg atoms, i.e., state \{1, 3\} corresponds to first and third atoms excited into Rydberg states. Their probabilities are represented by violet bars. From Fig.~\ref{Fig:MIS-Planar-Graphs}(a) we can see that the quantum states corresponding to solutions of the MIS demonstrate the highest probabilities. Finding the MISs of unit-disk graphs is a geometry-constrained problem, and the probabilities of MISs can be varied by adjusting the distance between vertices according to the phase diagram. The MISs of 3-Fan graph, shown in Fig.~\ref{Fig:MIS-Planar-Graphs}(b), are the same as in the $\mathds{P}_4$ graph, since vertex~1 is connected to all other vertices. However, the probabilities of the MISs in the 3-Fan graph are lower than the corresponding probabilities of the MISs in the $\mathds{P}_4$ graph. The sets \{1, 3\}, and \{2, 4\} of the $\mathds{P}_4$ and \{2, 4\} and \{3, 5\} of 3-Fan graph are of the same geometric pattern and consequently have the same probabilities. The error bars in the inset show the range of calculated errors for the MIS states (plotted in the same sequence as in the main figure) due to the fluctuations of the atomic positions. The white bars show the probability of a single atom being excited to the Rydberg state, which corresponds to an independent set of the graph.
%----%----%----%----%----%----%----%----%----%----%----%----%----%----%----%----%----%----%----%----%----%----%----%----%----%----%----%----%----%----%----%----
\begin{figure}[!]\centering
	\includegraphics[width=\columnwidth]{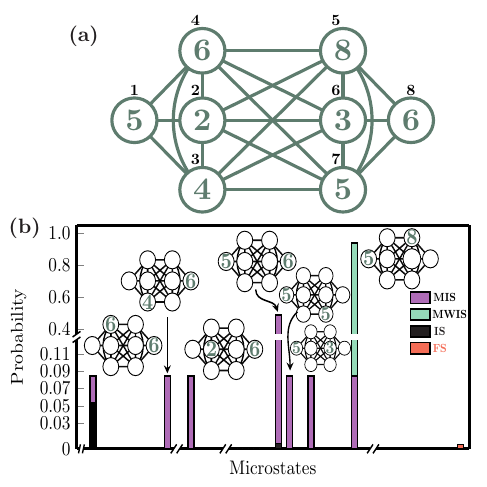}
	\caption{(a) 2D representation of the scheme of the Johnson solid $J_{14}$ graph. The black label is the vertex number and the green label represents the weight of each vertex. The exact positions of the vertices are given in Table~\ref{Table:AtomsPosition}. (b)Probabilities of MWISs for $J_{14}$. Some of the graph states with infinitely small probabilities are omitted from the plot.}
	\label{Fig:JohnsonGraphJ14}
\end{figure}
%----%----%----%----%----%----%----%----%----%----%----%----%----%----%----%----%----%----%----%----%----%----%----%----%----%----%----%----%----%----%----%----

The MISs of 3-Pan graph, shown in Fig.~\ref{Fig:MIS-Planar-Graphs}(c), are \textbf{\{}\{1, 4\}, \{2, 4\}\textbf{\}}. The corresponding states have almost equal probabilities, since both vertices~1and~2 are equally displaced from vertex 4. Also, the frustrated set or configuration \{1,2, 4\} appears in the calculations due to an imperfect Rydberg blockade for vertices~1 and~2. 

Fig.~\ref{Fig:MIS-Planar-Graphs}(d) shows the MISs of the Eiffel Tower graph [shown in Fig.~\ref{Fig:SysDynamics}(a)], which can be considered as combination of 3-Pan and 3-Fan graphs. MISs of the tower graph are \textbf{\{}\{1,5, 7\}, \{1,3, 5\}, \{1,4, 7\}, \{2,5, 7\}\textbf{\}}. Frustrated configurations are also present, but they have lower probabilities than those of the MISs. The MISs \textbf{\{}\{1,3, 5\},\{1,4, 7\}\textbf{\}} have identical geometric patterns and their probabilities are almost equal.

\subsection{Maximum-weight independent set \label{Sec:MWIS-PlanarGraphs}}

In this section we study the MWIS of the same graphs, which are shown in Fig.~\ref{Fig:SysDynamics}(a), but take into account the weights of their vertices. The positions of the vertices and weights are defined in Table~\ref{Table:AtomsPosition} and the weights are indicated in Fig.~\ref{Fig:SysDynamics}(a). The goal of the MWIS is to find the independent set with the maximum sum of its weights. The weight of each vertex is represented  by the maximum value of Rydberg detuning $\Delta_{f}$~\cite{byun2023rydberg}. The weights $\Delta_{f}$ of vertices are not equal. Therefore, the probability of exciting a particular atom to the Rydberg state is different. In Fig.~\ref{Fig:MWIS-Planar-Graphs} we show the calculated probabilities of transition of the atomic ensemble from the ground state $|g\rangle^{\otimes\tiny \text{N}}$ to  many-body states with a different configuration of Rydberg excitations. In Fig.~\ref{Fig:MWIS-Planar-Graphs}(a), the MWIS of the $\mathds{P}_4$ graph is the set \{1, 3\} which has a much higher probability  than other independent sets. Also, it can be noted that the probability of state \{2, 4\} is higher than that for \{1, 4\}, which is different from the results in Fig.~\ref{Fig:MIS-Planar-Graphs}(a) for the MIS problem. Analogous results are obtained in Fig.~\ref{Fig:MWIS-Planar-Graphs}(b). The MWIS for 3-Pan graph in Fig.~\ref{Fig:MWIS-Planar-Graphs}(c) is \{1, 4\}. It worth noting that  the calculated probability for the frustrated set \{1,2, 4\} is much higher than obtained before for the MIS problem for the same graph in Fig.\ref{Fig:MIS-Planar-Graphs}(c).

Overall, we note here that the MWIS problem is not only a geometry-constrained one as in the MIS. The probabilities of independent sets with the same geometric pattern, are not equal. Also, the probabilities of the MISs are sorted in ascending order of the sum of their weights.

%----%----%----%----%----%----%----%----%----%----%----%----%----%----%----%----%----%----%----%----%----%----%----%----%----%----%----%----%----%----%----%----%----%----%----%----%----%----%----%----%----%----%----%----%----%----%----%----%----%----%----%----%----%----%----%----%----%----%----%----%----%----%----%----%----%----%----%----%----%----%----%----%----%----%----%----%----%----%----%----%----%----%----%----%----%----%----%----%----%----%----%----%----%----%----%----%----%----%----%----%----%----%----%----%----%----%----%----%----%----%----%----%----%----%----%----%----%----%----%----%----%----%----%----%----%----%----%----%----%----%----%----%----%----%----%----%----%----%----%----%----%----%----%----%----%----%----%----%----%----%----%----%----%----%----%----%----%----%----%----%----%----%----%----%----%----%----%----

\section{Non-planar graph \label{Sec:NonPlanarGraph}}
%----%----%----%----%----%----%----%----%----%----%----%----%----%----%----%----%----%----%----%----%----%----%----%----%----%----%----%----%----%----%----%----

In this section we discuss the results of finding the MWIS and MIS of a nonplanar graph.  Fig.~\ref{Fig:JohnsonGraphJ14}(a) shows a 2D scheme representation of the $J_{14}$ graph. The $J_{14}$ is a graph with eight vertices in three dimensions, as defined in Table~\ref{Table:AtomsPosition}. The $J_{14}$ is known as \textit{elongated equilateral triangular bipyramid}, which is one of Johnson's 92 convex polyhedra solids~\cite{johnson1966convex}, whose facets are regular polygons. The graph includes $\mathcal{F}=9$ faces, as $6$ equilateral triangles, and $3$ squares. Euler's characteristic of this graph is $\chi=V-E+\mathcal{F}=2$. A similar nonplanar graph, represented in 2D, is called $\bar{X}_{152}$ as classified in the nomenclature of the Information System on Graph Classes and their Inclusions~\footnote{ \href{https://www.graphclasses.org/}{https://www.graphclasses.org/}}. The $J_{14}$ graph is a combination of two 3-Pan graphs.

In Fig.~\ref{Fig:JohnsonGraphJ14}(b) we plot the calculated probabilities of MISs and MWISs for the $J_{14}$ graph in two cases. \textbf{(i)} All vertices are equally weighted, which shows the MISs \textbf{\{}\{1, 5\}, \{1, 6\}, \{1, 7\}, \{1, 8\}, \{2, 8\}, \{3, 8\}, \{4, 8\}\textbf{\}} in violet-colored bars. The sets \textbf{\{}\{1, 5\}, \{1, 6\}, \{1, 7\}, \{2, 8\}, \{3, 8\}, \{4, 8\}\textbf{\}} have the same probability due to the fact that the atoms, representing vertices, are physically displaced equally from each other. Frustrated sets can be obtained from the simulation with probabilities much smaller than those of the MISs. \textbf{(ii)} The weights of vertices are not equal [the weight of each vertex is labeled in green in Fig.~\ref{Fig:JohnsonGraphJ14}(a)]. The MWIS according to the considered weights is \{1, 5\} as shown by a green bar in Fig.~\ref{Fig:JohnsonGraphJ14}(b). The independent sets, when considering the weighted graph, are the over plotted black bars. The frustrated set \{1,2,..., 8\} ($|r\rangle^{\otimes8}$) is shown by the red bar and exhibits infinitely low probability.

%----%----%----%----%----%----%----%----%----%----%----%----%----%----%----%----%----%----%----%----%----%----%----%----%----%----%----%----%----%----%----%----%----%----%----%----%----%----%----%----%----%----%----%----%----%----%----%----%----%----%----%----%----%----%----%----%----%----%----%----%----%----%----%----%----%----%----%----%----%----%----%----%----%----%----%----%----%----%----%----%----%----%----%----%----%----%----%----%----%----%----%----%----%----%----%----%----%----%----%----%----%----%----%----%----%----%----%----%----%----%----%----%----%----%----%----%----%----%----%----%----%----%----%----%----%----%----%----%----%----%----%----%----%----%----%----
\section{Quantum Wire \label{Sec:QuantumWire}}
%----%----%----%----%----%----%----%----%----%----%----%----%----%----%----%----%----%----%----%----%----%----%----%----%----%----%----%----%----%----%----%----
\begin{figure*}[!]\centering
	\includegraphics[width=\textwidth]{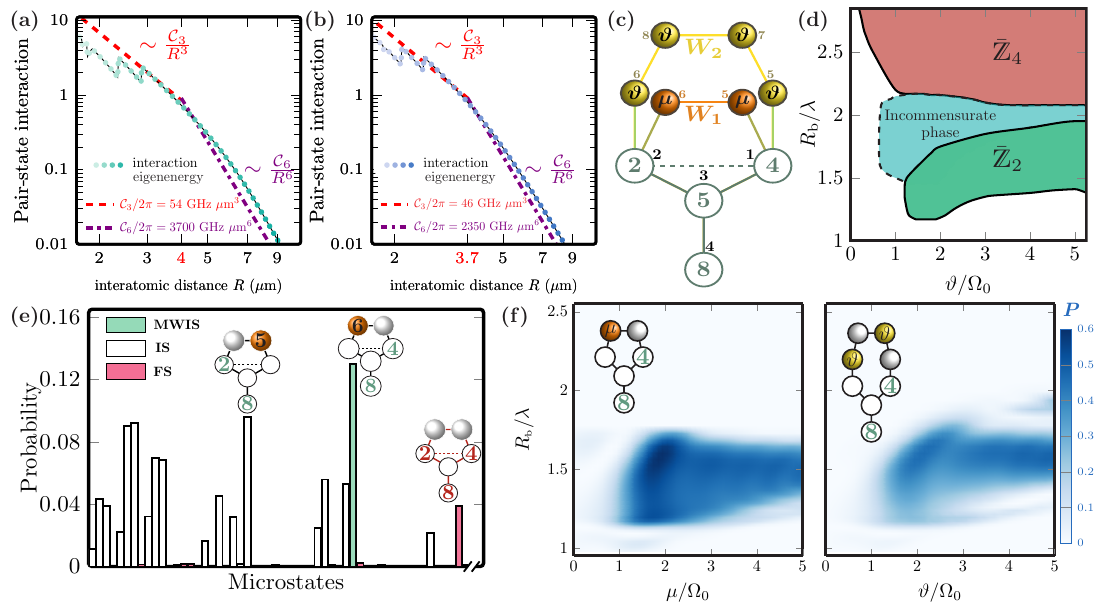}
	\caption{(a) The pair-state energy (\si{GHz}) for the heteronuclear interaction between Rb and Cs atoms excited simultaneously to Rydberg state $|81S_{1/2},m_j=1/2\rangle$ considering the same conditions and methods in Fig.~\ref{Fig:SysDynamics}(c). (b) The pair-state energy interaction energy between two Cs atoms. (c) The graph scheme of 3-Pan graph with vertices~1 and~2 are connected using two different spatial configurations of quantum wires $W_1$ (two vertices in violet color) and $W_2$ (four vertices in orange color). (d) The phase diagram of mediating the vertices $v_1$, and $v_2$ by the quantum wire $W_{2}$ which is composed of four equally weighted vertices (weight $\vartheta$~\si{MHz}). (e) The probabilities indicating the MWIS of 5-Pan graph (3-Pan+$W_1$). (f) The contour plots of the probability $\mathbf{P}$ of MWIS of 5-Pan (left panel) and 7-Pan (right panel) graphs as a function of dimensionless parameters $\mu/\Omega_{0}$ ($\vartheta/\Omega_{0}$) and $R_{\text{\tiny b}}/\lambda$.}
	\label{Fig:QWire}
\end{figure*}

%----%----%----%----%----%----%----%----%----%----%----%----%----%----%----%----%----%----%----%----%----%----%----%----%----%----%----%----%----%----%----%----

In this section, we  consider mediating strong interaction between distant vertices using the concept of a quantum wire~\cite{PRXQuantum.1.020311, kim2022rydberg}. Here we consider a dual-species quantum architecture by considering the wire to be composed of atoms from a different alkali-metal element used for graph representation. The dual-species architecture allows separation of the readout wavelengths during detection of the MIS or MWIS and suppression of crosstalk between neighboring atoms. The heteronuclear Rydberg interaction in atomic arrays was first discussed in Ref.~\cite{PhysRevA.92.042710}. The concept of dual-species quantum annealers was then introduced  in Ref.~\cite{glaetzle2017coherent}. A dual-species heteronuclear Rb-Cs array of ultracold atoms was experimentally demonstrated in Ref.~\cite{singh2022dual}. Schemes of controlled-NOT gates with several control and target atoms exploiting heteronuclear Rydberg interactions were studied in Ref.~\cite{Farouk2023parallel}. The two-qubit controlled-$Z$ gate with different chemical elements was demonstrated in Ref.~\cite{anand2024dual}, showing advantages compared with single-species homonuclear architectures. The scalability of  heteronuclear atomic architecture with coherent transport of control qubits was studied in Ref.~\cite{Farouk2023IJTP}. Analysis of heteronuclear interspecies interactions between Rydberg $d$-states of Rb and Cs atoms was presented in Ref.~\cite{ireland2024interspecies}. A dual-element model of quantum processors with single atoms or superatoms in the regime of Rydberg blockade was developed for quantum computations without the need of local addressing~\cite{FCesaQ2023}. Here we propose using a quantum wire of Cs atoms created from an array of traps generated by an acousto-optic deflector, with the graph represented by Rb atoms, which are loaded in the array of static traps generated by spatial light modulator. The ground state of Cs atoms of the quantum wire $|g\rangle_{\text{\tiny W}}=|6S_{1/2}, F=3, m_{F}=3\rangle$ are excited to the Rydberg state $|r\rangle_{\text{\tiny W}}=|81S_{1/2},m_j=1/2\rangle$ through the intermediate state $|m\rangle_{\text{\tiny W}}=|7P_{3/2},F=2\rangle$ by using $460$~\si{nm} and $1039$~\si{nm} laser lights, respectively. Atoms representing the graph and the wire are excited to the Rydberg state with the same principal quantum number~$n=81$. In this case the dominant regime of interaction is the van der Waals~\cite{browaeys2020many}.  Figs.~\ref{Fig:QWire}(a) and~\ref{Fig:QWire}(b) show the interaction between pairs of Rb and Cs atoms  (graph and wire) and pairs of Cs atoms (wire and wire), respectively. For an interatomic distance $R>R_{\text{\tiny vdW }} \simeq 4$~\si{\mu m}, it is guaranteed that all interactions are in the vdW regime. The values of $\mathcal{C}_3$ and $\mathcal{C}_6$ for all interactions are calculated by fitting the curves with the calculated interaction energy with all the parameters as in Fig.~\ref{Fig:LevelScheme}(c). The quantum wire should be in the anti-ferromagnetic phase, where at most only one of the adjacent atoms can be excited to the Rydberg state, known as the $\mathds{Z}_2$-phase.

Fig~\ref{Fig:QWire}(c) illustrates mediation of interaction between the vertices $v_1$ and $v_2$ of a 3-Pan graph by two quantum wires $W_1$ and $W_2$ formed by Cs atoms. The quantum wires are of different lengths (have different numbers of atoms) $\bar{w}_1=2$ and $\bar{w}_2=4$ . Atoms of each wire are equally weighted. The weights of the quantum wires are $\mu/2\pi$ for $W_1$ and $\vartheta/2\pi$ for $W_2$.

In Fig.~\ref{Fig:QWire}(d) we plot the phase diagram of a 3-Pan graph with quantum wire $W_2$ showing the probabilities as a function of two dimensionless parameters: the ratio of wire weights to the effective Rabi frequency $\vartheta/\Omega_{0}$, and the ratio of Rydberg blockade radius  to the graph spacing constant $R_{b}/\lambda$~\footnote{The Rydberg blockade radius $R_{b}$ of wire atoms is slightly different from the Rydberg blockade radius of graph atoms.}. In general, the obtained phase diagrams can vary for the same type of graph depending on the spatial arrangement of the atoms and the properties of the interactions between atoms. Hence, the discussion about phase diagrams here is valid only for the considered spatial arrangement of the atoms. If the quantum wire $W_2$, consisting of four atoms, is not connected to graph atoms, then the probability distribution of its quantum states for the same laser excitation pattern, as shown in Fig.~\ref{Fig:LevelScheme}(c), should behave similarly to the phase diagram, shown in Fig.~\ref{Fig:PhaseDiagramP4}. The phase diagram in Fig.~\ref{Fig:QWire}(d) shows the realization of a different ordered state $\bar{\mathds{Z}}_2$ [the bar sign over $\mathds{Z}_2$ indicates that the ordered states for a graph are different from the linear configuration from Fig.~\ref{Fig:SysDynamics}(d)]. The ordered state $\bar{\mathds{Z}}_4$ is also realized for the graph-wire states 
$|gggr\rangle_{\text{G}}|gggr\rangle_{\text{W}_2}$ and $|gggr\rangle_{\text{G}}|ggrg\rangle_{\text{W}_2}$.
%----%----%----%----%----%----%----%----%----%----%----%----%----%----%----%----%----%----%----%----%----%----%----%----%----%----%----%----%----%----%----%----
\begin{figure}[htb!]\centering
	\includegraphics[width=\columnwidth]{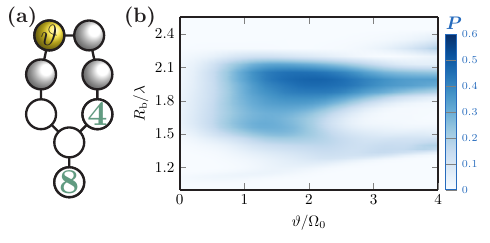}
	\caption{Probability distribution $\mathbf{P}$ of the incommensurate state $|rggr\rangle_{\text{G}}|gggr\rangle_{\text{W}_2}$.}
	\label{Fig:Incommensurate-Phases}
\end{figure}
%----%----%----%----%----%----%----%----%----%----%----%----%----%----%----%----%----%----%----%----%----%----%----%----%----%----%----%----%----%----%----%----

In Fig.~\ref{Fig:QWire}(e) we show the probabilities of finding the MWIS (\{1, 4\}=$|rggr\rangle_{G}|gr\rangle_{W_1}$) of the 3-Pan graph while using the quantum wire $W_1$ to mediate the interaction between $\text{v}_1$ and $\text{v}_2$. Finding the MWIS \{1, 4\} of the 3-Pan graph is independent of the wire weights $\mu$ as shown in Fig.~\ref{Fig:QWire}(f), and the MWIS is \{1,4, \textbf{6}\} (the bold text indicating the wire atom excited to the Rydberg state). We plot the probabilities of states $\{1,4, \textbf{6}\}= |rggr\rangle_{\scalebox{0.6}{G}} |gr\rangle_{\scalebox{0.6}{W}_1}$ and $\{1,4,\textbf{6}, \textbf{7}\} =|rggr \rangle_{ \scalebox{0.6}{G}} |grrg\rangle_{\scalebox{0.6}{W}_2}$ in Fig.~\ref{Fig:QWire}(f~i) and~\ref{Fig:QWire}(f~ii), respectively. The contour plot shows the probability of the considered state as a function of the weights of the wire atoms and distance parameter $\lambda$. In this figure we illustrate the effect of using quantum wires of different lengths (numbers of atoms). As shown, the minimum value of the vertex weight for the probability of the MWIS with a quantum wire $W_{1}$ formed by two auxiliary atoms is $\mu/\Omega_{0}\simeq0.92$ and for a quantum wire $W_{2}$ formed by four auxiliary atoms~$\vartheta/\Omega_{0}\simeq1.2$. The slightly increased cost for employing different lengths of a quantum wire to mediate the distant atoms is due to the spatial arrangement of the wire and graph atoms, which is interrupted by undesirable weak interactions.
%----%----%----%----%----%----%----%----%----%----%----%----%----%----%----%----%----%----%----%----%----%----%----%----%----%----%----%----%----%----%----%----
%----%----%----%----%----%----%----%----%----%----%----%----%----%----%----%----%----%----%----%----%----%----%----%----%----%----%----%----%----%----%----%----
%----%----%----%----%----%----%----%----%----%----%----%----%----%----%----%----%----%----%----%----%----%----%----%----%----%----%----%----%----%----%----%----

Due to the spatial arrangement of atoms representing the graph, an incommensurate phase emerges, showing a combination of $\bar{\mathds{Z}}_2$- and $\bar{\mathds{Z}}_3$-ordered states. The probability distribution of the incommensurate state is shown in Fig.~\ref{Fig:Incommensurate-Phases}. The realization of this incommensurate phase is controlled by the Rydberg detuning of each atom of the considered array in the 2D spatial arrangement of the atoms. This phase may boost realizing new dimer models which can be obtained from optimizing the interaction between atoms of homonuclear or heteronuclear architecture with the detuning of the Rydberg state of each atom. In Fig.~\ref{Fig:QWire}(f), the length of the quantum wire changes the dynamics of the system. In this case the quantum wire is not in an antiferromagnetic phase and that breaks the considered condition~\cite{kim2022rydberg}, then the quantum wire $W_2$ can be regarded as superatom~\cite{FCesaQ2023}.

In terms of phases of MWIS solutions, using the longer wire $W_2$ to find the solution is quite different from using the shorter wire $W_1$. In $W_1$ the only phase for the solution is the case when atom no. 6 is excited to the Rydberg state, which blocks Rydberg excitation of atoms no. 2 \& 5. At the same time, for the $W_2$ wire there are two phases, which is the case of Fig.~\ref{Fig:QWire}(f~ii) (atoms 6 and 7 are excited to the Rydberg state) in addition to Fig.~\ref{Fig:Incommensurate-Phases}(b) (only atom no. 8  is excited to Rydberg state). Both phases can be shown as solutions. The case shown in Fig.~\ref{Fig:Incommensurate-Phases}(b) is a solution with $\Delta_f/\Omega_0 \geq 1$, but for a different range of interatomic distances, compared with Fig.~\ref{Fig:QWire}(f~i).

\section{Conclusion\label{Sec:Conclusion}}

We proposed and optimized a quasi-adiabatic profile for sweeping the detuning of the Rydberg state and studied the quantum phase transitions from disordered to ordered-crystalline states of an 1D atomic array and realized $\mathds{Z}_2$- and $\mathds{Z}_3$-ordered states with the existence of a domain wall in  $\mathds{Z}_2$. Also, an incommensurate floating phase between $\mathds{Z}_3$ and $\mathds{Z}_4$ was obtained. We also investigated the effect of the sweeping rate on the probability of obtaining the $\mathds{Z}_2$-ordered state and calculated the corresponding critical detuning and studied the domain walls. Further optimization of the generation of quantum phases of matter can include special design of  composite laser pulses.

We considered solutions for the MIS and MWIS problems using arrays of ultracold neutral atoms excited to Rydberg states. We have obtained MISs and MWIS of planar and nonplanar graphs. For the MIS problem, we concluded that it is a geometry-constrained problem, and generation of the independent sets for the same geometric patterns  have equal probabilities. For the MWIS problem the weights of each vertex change significantly the character of the resulting most probable independent set. The probabilities of MISs of the weighted graph are sort in descending order by the sum weights of each set. The use of quantum wire can help mediate strong interactions of distant vertices. Also, quantum wires can be used to perform quantum annealing of non unit-disk graphs. Using the heteronuclear structure of the atomic array it is feasible to distinguish the measurement of atomic states representing the graph and the wire vertices due to the separation of the wavelengths and reduction of cross-talk among different chemical elements. Moreover, the cost $\Delta_{f}$ of finding the MIS or MWIS increases proportionally for longer lengths of quantum wires.

An incommensurate floating phase between $\bar{\mathds{Z}}_2$ and $\bar{\mathds{Z}}_3$ in the 2D atomic array, formed as 7-Pan graph representation, was realized, allowing us to use a quantum wire, which is not in the antiferromagnetic state, for a solution of the MWIS. This incommensurate phase does not exist in 1D arrays for equal detunings of transitions to the Rydberg state. The results of incommensurate phases can open new directions for realizing exotic states of matter.

\textit{Note added:} During the completion of this manuscript we became aware of the foremost experimental demonstration of the weighted graphs, verifying the ability to prepare weighted graphs in 1D and 2D arrays~\cite{de2024demonstration}.

%----%----%----%----%----%----%----%----%----%----%----%----%----%----%----%----%----%----%----%----%----%----%----%----%----%----%----%----%----%----%----%----%----%----%----%----%----%----%----%----%----%----%----%----%----%----%----%----%----%----%----%----%----%----%----%----%----%----%----%----%----%----%----%----%----%----%----%----%----%----%----%----%----%----%----%----%----%----%----%----%----%----%----%----%----%----%----%----%----%----%----%----%----%----%----%----%----%----%----%----%----%----%----%----%----%----%----%----%----%----%----%----%----%----%----%----%----%----%----%----%----%----%----%----%----%----%----%----%----%----%----%----%----%----%----%----%----%----%----%----%----%----%----%----%----%----%----%----%----%----%----%----%----%----%----%----%----%----%----%----%----%----%----%----%----%----%----%----
%----%----%----%----%----%----%----%----%----%----%----%----%----%----%----%----%----%----%----%----%----%----%----%----%----%----%----%----%----%----%----%----%----%----%----%----%----%----%----%----%----%----%----%----%----%----%----%----%----%----%----%----%----%----%----%----%----%----%----%----%----%----%----%----%----%----%----%----%----%----%----%----%----%----%----%----%----%----%----%----
\begin{table*}[!htb]\centering
	\caption{Sites of atoms demonstrating the considered graphs as a function of graph spacing constant $\lambda$.}
	\label{Table:AtomsPosition}
	\normalfont\addtolength{\tabcolsep}{0 pt}
	\def\arraystretch{1.0}
	\begin{tabular}{|c|llll|}
		\toprule
		Graph & \multicolumn{4}{l|}{Positions} \\ \hline
		%----%----%----%----%----%----%----%----%----%----%
		%----%----%----%----%----%----%----%----%----%----%
		%----%----%----%----%----%----%----%----%----%----%
		$\mathds{P}_4$ & \textbf{1:}~$\left( -\lambda, 0 , 0 \right)$ & \textbf{2:}~$\left(-\frac{1}{2} \lambda,\frac{\sqrt{3}}{2} \lambda,0\right)$ & \textbf{3:}~$\left(\frac{1}{2} \lambda,\frac{\sqrt{3}}{2} \lambda,0\right)$ &\\ 
		$\left[\text{Fig.}~\ref{Fig:SysDynamics}(a)\right]$ & \multicolumn{4}{l|}{\textbf{4:}~$\left( \lambda, 0 , 0 \right)$}
		\\ \hline
		%----%----%----%----%----%----%----%----%----%----%
		%----%----%----%----%----%----%----%----%----%----%
		%----%----%----%----%----%----%----%----%----%----%
		3-Fan & \textbf{1:}~$\left( 0, 0 , 0 \right)$ &\textbf{2:}~$\left( -\lambda, 0 , 0 \right)$ & \textbf{3:}~$\left(-\frac{1}{2} \lambda,\frac{\sqrt{3}}{2} \lambda,0\right)$ &\\ 
		$\left[\text{Fig.}~\ref{Fig:SysDynamics}(a)\right]$ & \textbf{4:}~$\left(\frac{1}{2} \lambda,\frac{\sqrt{3}}{2} \lambda,0\right)$ & \textbf{5:}~$\left( \lambda, 0 , 0 \right)$ & &\\ \hline
		%----%----%----%----%----%----%----%----%----%----%
		%----%----%----%----%----%----%----%----%----%----%
		%----%----%----%----%----%----%----%----%----%----%
		3-Pan & \textbf{1:}~($\frac{-\lambda}{\sqrt{2}}$, $\frac{\lambda}{\sqrt{2}}$, 0) &
		\textbf{2:}~($\frac{-\lambda}{\sqrt{2}}$, $\frac{-\lambda}{\sqrt{2}}$, 0) & \textbf{3:}~(0, 0, 0) &\\ $\left[\text{Fig.}~\ref{Fig:SysDynamics}(a)\right]$ &
		\textbf{4:}~($\lambda$, 0, 0) &&&
		\\
		---------&\multicolumn{4}{c|}{--------------------------------------------------------------------------------------------------------------------------------} \\
		5-Pan ($W_1$)& \multicolumn{2}{l}{\textbf{5:}~$\left( -\frac{1}{\sqrt{2}}\lambda-\frac{1}{2}\lambda \sqrt{1+2\sqrt{2}}, \frac{1}{2}\lambda,0 \right)$} & \multicolumn{2}{l|}{\textbf{6:}~$\left( -\frac{1}{\sqrt{2}}\lambda-\frac{1}{2}\lambda \sqrt{1+2\sqrt{2}}, -\frac{1}{2}\lambda, 0 \right)$}
		\\	$\left[\text{Fig.}~\ref{Fig:QWire}(c)\right]$ &&&&\\
		---------&\multicolumn{4}{c|}{--------------------------------------------------------------------------------------------------------------------------------} \\
		7-Pan ($W_2$)& \multicolumn{2}{l}{\textbf{5:}~$\left( -\frac{1}{\sqrt{2}}\lambda-\lambda, \frac{1}{\sqrt{2}}\lambda,0 \right)$}
		&
		\multicolumn{2}{l|}{\textbf{6:}~$\left( -\frac{1}{\sqrt{2}}\lambda-\lambda, -\frac{1}{\sqrt{2}}\lambda,0 \right)$}
		\\$\left[\text{Fig.}~\ref{Fig:QWire}(c)\right]$&
		\multicolumn{2}{l}{\textbf{7:}~$\left( -\frac{1}{2}(2+\sqrt{2}+\sqrt{1+2\sqrt{2}})\lambda , \frac{1}{2}\lambda,0 \right)$}
		&
		\multicolumn{2}{l|}{\textbf{8:}~$\left( -\frac{1}{2}(2+\sqrt{2}+\sqrt{1+2\sqrt{2}})\lambda ,- \frac{1}{2}\lambda,0 \right)$}\\
		\hline
		%----%----%----%----%----%----%----%----%----%----%
		%----%----%----%----%----%----%----%----%----%----%
		%----%----%----%----%----%----%----%----%----%----%
		Tower & \textbf{1:}~$\left(0, 0, 0\right)$,&
		\textbf{2:}~$\left(0, -\lambda, 0\right)$,&
		\textbf{3:}~$\left(\frac{-\lambda}{\sqrt{2}}~\frac{-\lambda}{\sqrt{2}}-\lambda, 0\right)$
		& \\ 
		$\left[\text{Fig.}~\ref{Fig:SysDynamics}(a)\right]$& 
		\textbf{4:}~$\left( \frac{\lambda}{\sqrt{2}}, \frac{-\lambda}{\sqrt{2}}-\lambda, 0 \right)$ & \textbf{5:}~$\left(\frac{2\lambda}{\sqrt{2}}, \frac{-2\lambda}{\sqrt{2}}-\lambda, 0 \right)$&
		\textbf{6:}~$\left( 0, -\lambda(1+\sqrt{2}), 0 \right)$&
		\\&
		\multicolumn{4}{l|}{\textbf{7:}~$\left(\frac{-2\lambda}{\sqrt{2}}, \frac{-2\lambda}{\sqrt{2}}-\lambda, 0\right)$} \\ \hline
		%----%----%----%----%----%----%----%----%----%----%
		%----%----%----%----%----%----%----%----%----%----%
		%----%----%----%----%----%----%----%----%----%----%
		$J_{14}/\bar{X}_{152}$ & \textbf{1:}~$\left(x_0, y_0, \eta_{0} \lambda\right)$
		& \textbf{2:}~$\left(x_0, y_0 + \frac{\lambda}{\sqrt{3}} , -\lambda\right)$ & \textbf{3:}~$\left( x_0 - \frac{\lambda}{2}, y_0 - \frac{\lambda}{2\sqrt{3}}, -\lambda \right)$&
		\\
		$\left[\text{Fig.}~\ref{Fig:JohnsonGraphJ14}(a)\right]$&
		\textbf{4:}~$\left( x_0 + \frac{\lambda}{2}, y_0 - \frac{\lambda}{2\sqrt{3}}, -\lambda \right)$ & \textbf{5:}~$\left( x_0 + \frac{\lambda}{2}, y_0 - \frac{\lambda}{2\sqrt{3}}, -2\lambda*q_0 \right)$
		& \textbf{6:}~$\left( x_0 , y_0 + \frac{\lambda}{\sqrt{3}}, -2\lambda*q_0 \right)$ &
		\\&
		\multicolumn{2}{l}{\textbf{7:}~$\bigl( x_0 - \frac{\lambda}{2}, y_0 - \frac{\lambda}{2\sqrt{3}}, -2\lambda*q_0 \bigl)$} & \multicolumn{2}{l|}{\textbf{8:}~$\bigl(x_0, y_0 , -\lambda(\eta_{0} + 2q_0 +  1 ) \bigl)$}
		\\ & 
		\multicolumn{4}{l|}{where $x_0$, and $y_0$ are arbitrary points. In calculations, we considered $x_0=y_0=0$.} \\ &\multicolumn{4}{l|}{To get equal lengths of triangle and square sides, we set $\eta_0 \simeq -0.183368$, and $q_0=1$.}
		\\
		\toprule
		%\hline
	\end{tabular}
\end{table*}
%----%----%----%----%----%----%----%----%----%----%----%----%----%----%----%----%----%----%----%----%----%----%----%----%----%----%----%----%----%----%----%----%----%----%----%----%----%----%----%----%----%----%----%----%----%----%----%----%----%----%----%----%----%----%----%----%----%----%----%----%----%----%----

\begin{acknowledgments}
This work was supported by the Russian Science Foundation (Grant No.~\href{https://rscf.ru/project/23-42-00031/}{23-42-00031}). A. M. Farouk acknowledges financial support from the joint executive educational program between Egypt and Russia (Grant No. EGY-6544/19). P. X. acknowledges financial support from the National Key Research and Development Program of China (Grant No. 2021YFA1402001) and the National Natural Science Foundation of China (Grants No. 12261131507 and No. U20A2074). Data sets supporting plots in this paper are available upon request from the corresponding author.

\end{acknowledgments}

%----%----%----%----%----%----%----%----%----%----%----%----%----%----%----%----%----%----%----%----%----%----%----%----%----%----%----%----%----%----%----%----%----%----%----%----%----%----%----%----%----%----%----%----%----%----%----%----%----%----%----%----%----%----%----%----%----%----%----%----%----%----%----
%\appendix

%\newpage

\nocite{*}

\bibliography{References}% Produces the bibliography via BibTeX.

\begin{thebibliography}{69}%
\makeatletter
\providecommand \@ifxundefined [1]{%
 \@ifx{#1\undefined}
}%
\providecommand \@ifnum [1]{%
 \ifnum #1\expandafter \@firstoftwo
 \else \expandafter \@secondoftwo
 \fi
}%
\providecommand \@ifx [1]{%
 \ifx #1\expandafter \@firstoftwo
 \else \expandafter \@secondoftwo
 \fi
}%
\providecommand \natexlab [1]{#1}%
\providecommand \enquote  [1]{``#1''}%
\providecommand \bibnamefont  [1]{#1}%
\providecommand \bibfnamefont [1]{#1}%
\providecommand \citenamefont [1]{#1}%
\providecommand \href@noop [0]{\@secondoftwo}%
\providecommand \href [0]{\begingroup \@sanitize@url \@href}%
\providecommand \@href[1]{\@@startlink{#1}\@@href}%
\providecommand \@@href[1]{\endgroup#1\@@endlink}%
\providecommand \@sanitize@url [0]{\catcode `\\12\catcode `\$12\catcode
  `\&12\catcode `\#12\catcode `\^12\catcode `\_12\catcode `\%12\relax}%
\providecommand \@@startlink[1]{}%
\providecommand \@@endlink[0]{}%
\providecommand \url  [0]{\begingroup\@sanitize@url \@url }%
\providecommand \@url [1]{\endgroup\@href {#1}{\urlprefix }}%
\providecommand \urlprefix  [0]{URL }%
\providecommand \Eprint [0]{\href }%
\providecommand \doibase [0]{https://doi.org/}%
\providecommand \selectlanguage [0]{\@gobble}%
\providecommand \bibinfo  [0]{\@secondoftwo}%
\providecommand \bibfield  [0]{\@secondoftwo}%
\providecommand \translation [1]{[#1]}%
\providecommand \BibitemOpen [0]{}%
\providecommand \bibitemStop [0]{}%
\providecommand \bibitemNoStop [0]{.\EOS\space}%
\providecommand \EOS [0]{\spacefactor3000\relax}%
\providecommand \BibitemShut  [1]{\csname bibitem#1\endcsname}%
\let\auto@bib@innerbib\@empty
%</preamble>
\bibitem [{\citenamefont {Arora}\ and\ \citenamefont
  {Barak}(2009)}]{arora2009computational}%
  \BibitemOpen
  \bibfield  {author} {\bibinfo {author} {\bibfnamefont {S.}~\bibnamefont
  {Arora}}\ and\ \bibinfo {author} {\bibfnamefont {B.}~\bibnamefont {Barak}},\
  }\href@noop {} {\emph {\bibinfo {title} {{Computational Complexity: A Modern
  Approach}}}}\ (\bibinfo  {publisher} {Cambridge University Press,
  Cambridge},\ \bibinfo {year} {2009})\BibitemShut {NoStop}%
\bibitem [{\citenamefont {Wagner}\ \emph {et~al.}(2024)\citenamefont {Wagner},
  \citenamefont {Poole}, \citenamefont {Graham},\ and\ \citenamefont
  {Saffman}}]{wagner2023benchmarking}%
  \BibitemOpen
  \bibfield  {author} {\bibinfo {author} {\bibfnamefont {N.}~\bibnamefont
  {Wagner}}, \bibinfo {author} {\bibfnamefont {C.}~\bibnamefont {Poole}},
  \bibinfo {author} {\bibfnamefont {T.}~\bibnamefont {Graham}},\ and\ \bibinfo
  {author} {\bibfnamefont {M.}~\bibnamefont {Saffman}},\ }\bibfield  {title}
  {\bibinfo {title} {{Benchmarking a neutral-atom quantum computer}},\ }\href
  {https://doi.org/10.1142/S0219749924500011} {\bibfield  {journal} {\bibinfo
  {journal} {International Journal of Quantum Information}\ }\textbf {\bibinfo
  {volume} {22}},\ \bibinfo {pages} {2450001} (\bibinfo {year}
  {2024})}\BibitemShut {NoStop}%
\bibitem [{\citenamefont {Kim}\ \emph {et~al.}(2022)\citenamefont {Kim},
  \citenamefont {Kim}, \citenamefont {Hwang}, \citenamefont {Moon},\ and\
  \citenamefont {Ahn}}]{kim2022rydberg}%
  \BibitemOpen
  \bibfield  {author} {\bibinfo {author} {\bibfnamefont {M.}~\bibnamefont
  {Kim}}, \bibinfo {author} {\bibfnamefont {K.}~\bibnamefont {Kim}}, \bibinfo
  {author} {\bibfnamefont {J.}~\bibnamefont {Hwang}}, \bibinfo {author}
  {\bibfnamefont {E.-G.}\ \bibnamefont {Moon}},\ and\ \bibinfo {author}
  {\bibfnamefont {J.}~\bibnamefont {Ahn}},\ }\bibfield  {title} {\bibinfo
  {title} {{Rydberg quantum wires for maximum independent set problems}},\
  }\href {https://doi.org/10.1038/s41567-022-01629-5} {\bibfield  {journal}
  {\bibinfo  {journal} {Nature Physics}\ }\textbf {\bibinfo {volume} {18}},\
  \bibinfo {pages} {755} (\bibinfo {year} {2022})}\BibitemShut {NoStop}%
\bibitem [{\citenamefont {Glaetzle}\ \emph {et~al.}(2017)\citenamefont
  {Glaetzle}, \citenamefont {van Bijnen}, \citenamefont {Zoller},\ and\
  \citenamefont {Lechner}}]{glaetzle2017coherent}%
  \BibitemOpen
  \bibfield  {author} {\bibinfo {author} {\bibfnamefont {A.~W.}\ \bibnamefont
  {Glaetzle}}, \bibinfo {author} {\bibfnamefont {R.~M.}\ \bibnamefont {van
  Bijnen}}, \bibinfo {author} {\bibfnamefont {P.}~\bibnamefont {Zoller}},\ and\
  \bibinfo {author} {\bibfnamefont {W.}~\bibnamefont {Lechner}},\ }\bibfield
  {title} {\bibinfo {title} {{A coherent quantum annealer with Rydberg
  atoms}},\ }\href {https://doi.org/10.1038/ncomms15813} {\bibfield  {journal}
  {\bibinfo  {journal} {Nature communications}\ }\textbf {\bibinfo {volume}
  {8}},\ \bibinfo {pages} {15813} (\bibinfo {year} {2017})}\BibitemShut
  {NoStop}%
\bibitem [{\citenamefont {Bondy}\ and\ \citenamefont
  {Murty}(1982)}]{bondy1982graph}%
  \BibitemOpen
  \bibfield  {author} {\bibinfo {author} {\bibfnamefont {J.~A.}\ \bibnamefont
  {Bondy}}\ and\ \bibinfo {author} {\bibfnamefont {U.~S.~R.}\ \bibnamefont
  {Murty}},\ }\href@noop {} {\emph {\bibinfo {title} {{Graph Theory with
  Applications}}}}\ (\bibinfo  {publisher} {North-Holland, Amsterdam},\
  \bibinfo {year} {1982})\BibitemShut {NoStop}%
\bibitem [{\citenamefont {Graham}\ \emph {et~al.}(2022)\citenamefont {Graham},
  \citenamefont {Song}, \citenamefont {Scott}, \citenamefont {Poole},
  \citenamefont {Phuttitarn}, \citenamefont {Jooya}, \citenamefont {Eichler},
  \citenamefont {Jiang}, \citenamefont {Marra}, \citenamefont {Grinkemeyer}
  \emph {et~al.}}]{graham2022multi}%
  \BibitemOpen
  \bibfield  {author} {\bibinfo {author} {\bibfnamefont {T.}~\bibnamefont
  {Graham}}, \bibinfo {author} {\bibfnamefont {Y.}~\bibnamefont {Song}},
  \bibinfo {author} {\bibfnamefont {J.}~\bibnamefont {Scott}}, \bibinfo
  {author} {\bibfnamefont {C.}~\bibnamefont {Poole}}, \bibinfo {author}
  {\bibfnamefont {L.}~\bibnamefont {Phuttitarn}}, \bibinfo {author}
  {\bibfnamefont {K.}~\bibnamefont {Jooya}}, \bibinfo {author} {\bibfnamefont
  {P.}~\bibnamefont {Eichler}}, \bibinfo {author} {\bibfnamefont
  {X.}~\bibnamefont {Jiang}}, \bibinfo {author} {\bibfnamefont
  {A.}~\bibnamefont {Marra}}, \bibinfo {author} {\bibfnamefont
  {B.}~\bibnamefont {Grinkemeyer}}, \emph {et~al.},\ }\bibfield  {title}
  {\bibinfo {title} {{Multi-qubit entanglement and algorithms on a neutral-atom
  quantum computer}},\ }\href {https://doi.org/10.1038/s41586-022-04603-6}
  {\bibfield  {journal} {\bibinfo  {journal} {Nature}\ }\textbf {\bibinfo
  {volume} {604}},\ \bibinfo {pages} {457} (\bibinfo {year}
  {2022})}\BibitemShut {NoStop}%
\bibitem [{\citenamefont {Zhou}\ \emph {et~al.}(2020)\citenamefont {Zhou},
  \citenamefont {Wang}, \citenamefont {Choi}, \citenamefont {Pichler},\ and\
  \citenamefont {Lukin}}]{zhou2020quantum}%
  \BibitemOpen
  \bibfield  {author} {\bibinfo {author} {\bibfnamefont {L.}~\bibnamefont
  {Zhou}}, \bibinfo {author} {\bibfnamefont {S.-T.}\ \bibnamefont {Wang}},
  \bibinfo {author} {\bibfnamefont {S.}~\bibnamefont {Choi}}, \bibinfo {author}
  {\bibfnamefont {H.}~\bibnamefont {Pichler}},\ and\ \bibinfo {author}
  {\bibfnamefont {M.~D.}\ \bibnamefont {Lukin}},\ }\bibfield  {title} {\bibinfo
  {title} {{Quantum approximate optimization algorithm: Performance, mechanism,
  and implementation on near-term devices}},\ }\href
  {https://doi.org/10.1103/PhysRevX.10.021067} {\bibfield  {journal} {\bibinfo
  {journal} {Physical Review X}\ }\textbf {\bibinfo {volume} {10}},\ \bibinfo
  {pages} {021067} (\bibinfo {year} {2020})}\BibitemShut {NoStop}%
\bibitem [{\citenamefont {Mu\~noz Arias}\ \emph {et~al.}(2024)\citenamefont
  {Mu\~noz Arias}, \citenamefont {Kourtis},\ and\ \citenamefont
  {Blais}}]{munoz2023low}%
  \BibitemOpen
  \bibfield  {author} {\bibinfo {author} {\bibfnamefont {M.~H.}\ \bibnamefont
  {Mu\~noz Arias}}, \bibinfo {author} {\bibfnamefont {S.}~\bibnamefont
  {Kourtis}},\ and\ \bibinfo {author} {\bibfnamefont {A.}~\bibnamefont
  {Blais}},\ }\bibfield  {title} {\bibinfo {title} {{Low-depth Clifford
  circuits approximately solve MaxCut}},\ }\href
  {https://doi.org/10.1103/PhysRevResearch.6.023294} {\bibfield  {journal}
  {\bibinfo  {journal} {Phys. Rev. Res.}\ }\textbf {\bibinfo {volume} {6}},\
  \bibinfo {pages} {023294} (\bibinfo {year} {2024})}\BibitemShut {NoStop}%
\bibitem [{\citenamefont {Paradezhenko}\ \emph {et~al.}(2024)\citenamefont
  {Paradezhenko}, \citenamefont {Pervishko},\ and\ \citenamefont
  {Yudin}}]{paradezhenko2023probabilistic}%
  \BibitemOpen
  \bibfield  {author} {\bibinfo {author} {\bibfnamefont {G.~V.}\ \bibnamefont
  {Paradezhenko}}, \bibinfo {author} {\bibfnamefont {A.~A.}\ \bibnamefont
  {Pervishko}},\ and\ \bibinfo {author} {\bibfnamefont {D.}~\bibnamefont
  {Yudin}},\ }\bibfield  {title} {\bibinfo {title} {{Probabilistic tensor
  optimization of quantum circuits for the max-$k$-cut problem}},\ }\href
  {https://doi.org/10.1103/PhysRevA.109.012436} {\bibfield  {journal} {\bibinfo
   {journal} {Phys. Rev. A}\ }\textbf {\bibinfo {volume} {109}},\ \bibinfo
  {pages} {012436} (\bibinfo {year} {2024})}\BibitemShut {NoStop}%
\bibitem [{\citenamefont {Brady}\ and\ \citenamefont
  {Hadfield}(2023)}]{brady2023iterative}%
  \BibitemOpen
  \bibfield  {author} {\bibinfo {author} {\bibfnamefont {L.~T.}\ \bibnamefont
  {Brady}}\ and\ \bibinfo {author} {\bibfnamefont {S.}~\bibnamefont
  {Hadfield}},\ }\bibfield  {title} {\bibinfo {title} {{Iterative Quantum
  Algorithms for Maximum Independent Set: A Tale of Low-Depth Quantum
  Algorithms}},\ }\bibfield  {journal} {\bibinfo  {journal} {arXiv preprint
  arXiv:2309.13110}\ }\href {https://doi.org/10.48550/arXiv.2309.13110}
  {10.48550/arXiv.2309.13110} (\bibinfo {year} {2023})\BibitemShut {NoStop}%
\bibitem [{\citenamefont {Fin{\v{z}}gar}\ \emph {et~al.}(2024)\citenamefont
  {Fin{\v{z}}gar}, \citenamefont {Kerschbaumer}, \citenamefont {Schuetz},
  \citenamefont {Mendl},\ and\ \citenamefont
  {Katzgraber}}]{finvzgar2024quantum}%
  \BibitemOpen
  \bibfield  {author} {\bibinfo {author} {\bibfnamefont {J.~R.}\ \bibnamefont
  {Fin{\v{z}}gar}}, \bibinfo {author} {\bibfnamefont {A.}~\bibnamefont
  {Kerschbaumer}}, \bibinfo {author} {\bibfnamefont {M.~J.}\ \bibnamefont
  {Schuetz}}, \bibinfo {author} {\bibfnamefont {C.~B.}\ \bibnamefont {Mendl}},\
  and\ \bibinfo {author} {\bibfnamefont {H.~G.}\ \bibnamefont {Katzgraber}},\
  }\bibfield  {title} {\bibinfo {title} {{Quantum-informed recursive
  optimization algorithms}},\ }\href
  {https://doi.org/10.1103/PRXQuantum.5.020327} {\bibfield  {journal} {\bibinfo
   {journal} {PRX Quantum}\ }\textbf {\bibinfo {volume} {5}},\ \bibinfo {pages}
  {020327} (\bibinfo {year} {2024})}\BibitemShut {NoStop}%
\bibitem [{\citenamefont {Bauer}\ \emph {et~al.}(2024)\citenamefont {Bauer},
  \citenamefont {Yeter-Aydeniz}, \citenamefont {Kokkas},\ and\ \citenamefont
  {Siopsis}}]{bauer2024solving}%
  \BibitemOpen
  \bibfield  {author} {\bibinfo {author} {\bibfnamefont {N.}~\bibnamefont
  {Bauer}}, \bibinfo {author} {\bibfnamefont {K.}~\bibnamefont
  {Yeter-Aydeniz}}, \bibinfo {author} {\bibfnamefont {E.}~\bibnamefont
  {Kokkas}},\ and\ \bibinfo {author} {\bibfnamefont {G.}~\bibnamefont
  {Siopsis}},\ }\bibfield  {title} {\bibinfo {title} {{Solving Power Grid
  Optimization Problems with Rydberg Atoms}},\ }\bibfield  {journal} {\bibinfo
  {journal} {arXiv preprint arXiv:2404.11440}\ }\href
  {https://doi.org/10.48550/arXiv.2404.11440} {10.48550/arXiv.2404.11440}
  (\bibinfo {year} {2024})\BibitemShut {NoStop}%
\bibitem [{\citenamefont {Farhi}\ \emph {et~al.}(2000)\citenamefont {Farhi},
  \citenamefont {Goldstone}, \citenamefont {Gutmann},\ and\ \citenamefont
  {Sipser}}]{farhi2000quantum}%
  \BibitemOpen
  \bibfield  {author} {\bibinfo {author} {\bibfnamefont {E.}~\bibnamefont
  {Farhi}}, \bibinfo {author} {\bibfnamefont {J.}~\bibnamefont {Goldstone}},
  \bibinfo {author} {\bibfnamefont {S.}~\bibnamefont {Gutmann}},\ and\ \bibinfo
  {author} {\bibfnamefont {M.}~\bibnamefont {Sipser}},\ }\bibfield  {title}
  {\bibinfo {title} {{Quantum computation by adiabatic evolution}},\ }\bibfield
   {journal} {\bibinfo  {journal} {arXiv preprint quant-ph/0001106}\ }\href
  {https://doi.org/10.48550/arXiv.quant-ph/0001106}
  {10.48550/arXiv.quant-ph/0001106} (\bibinfo {year} {2000})\BibitemShut
  {NoStop}%
\bibitem [{\citenamefont {Schiffer}\ \emph {et~al.}(2022)\citenamefont
  {Schiffer}, \citenamefont {Tura},\ and\ \citenamefont
  {Cirac}}]{PRXQuantum.3.020347}%
  \BibitemOpen
  \bibfield  {author} {\bibinfo {author} {\bibfnamefont {B.~F.}\ \bibnamefont
  {Schiffer}}, \bibinfo {author} {\bibfnamefont {J.}~\bibnamefont {Tura}},\
  and\ \bibinfo {author} {\bibfnamefont {J.~I.}\ \bibnamefont {Cirac}},\
  }\bibfield  {title} {\bibinfo {title} {{Adiabatic Spectroscopy and a
  Variational Quantum Adiabatic Algorithm}},\ }\href
  {https://doi.org/10.1103/PRXQuantum.3.020347} {\bibfield  {journal} {\bibinfo
   {journal} {PRX Quantum}\ }\textbf {\bibinfo {volume} {3}},\ \bibinfo {pages}
  {020347} (\bibinfo {year} {2022})}\BibitemShut {NoStop}%
\bibitem [{\citenamefont {Scholl}\ \emph {et~al.}(2021)\citenamefont {Scholl},
  \citenamefont {Schuler}, \citenamefont {Williams}, \citenamefont
  {Eberharter}, \citenamefont {Barredo}, \citenamefont {Schymik}, \citenamefont
  {Lienhard}, \citenamefont {Henry}, \citenamefont {Lang}, \citenamefont
  {Lahaye} \emph {et~al.}}]{scholl2021quantum}%
  \BibitemOpen
  \bibfield  {author} {\bibinfo {author} {\bibfnamefont {P.}~\bibnamefont
  {Scholl}}, \bibinfo {author} {\bibfnamefont {M.}~\bibnamefont {Schuler}},
  \bibinfo {author} {\bibfnamefont {H.~J.}\ \bibnamefont {Williams}}, \bibinfo
  {author} {\bibfnamefont {A.~A.}\ \bibnamefont {Eberharter}}, \bibinfo
  {author} {\bibfnamefont {D.}~\bibnamefont {Barredo}}, \bibinfo {author}
  {\bibfnamefont {K.-N.}\ \bibnamefont {Schymik}}, \bibinfo {author}
  {\bibfnamefont {V.}~\bibnamefont {Lienhard}}, \bibinfo {author}
  {\bibfnamefont {L.-P.}\ \bibnamefont {Henry}}, \bibinfo {author}
  {\bibfnamefont {T.~C.}\ \bibnamefont {Lang}}, \bibinfo {author}
  {\bibfnamefont {T.}~\bibnamefont {Lahaye}}, \emph {et~al.},\ }\bibfield
  {title} {\bibinfo {title} {{Quantum simulation of 2D antiferromagnets with
  hundreds of Rydberg atoms}},\ }\href
  {https://doi.org/10.1038/s41586-021-03585-1} {\bibfield  {journal} {\bibinfo
  {journal} {Nature}\ }\textbf {\bibinfo {volume} {595}},\ \bibinfo {pages}
  {233} (\bibinfo {year} {2021})}\BibitemShut {NoStop}%
\bibitem [{\citenamefont {Ebadi}\ \emph {et~al.}(2021)\citenamefont {Ebadi},
  \citenamefont {Wang}, \citenamefont {Levine}, \citenamefont {Keesling},
  \citenamefont {Semeghini}, \citenamefont {Omran}, \citenamefont {Bluvstein},
  \citenamefont {Samajdar}, \citenamefont {Pichler}, \citenamefont {Ho} \emph
  {et~al.}}]{ebadi2021quantum}%
  \BibitemOpen
  \bibfield  {author} {\bibinfo {author} {\bibfnamefont {S.}~\bibnamefont
  {Ebadi}}, \bibinfo {author} {\bibfnamefont {T.~T.}\ \bibnamefont {Wang}},
  \bibinfo {author} {\bibfnamefont {H.}~\bibnamefont {Levine}}, \bibinfo
  {author} {\bibfnamefont {A.}~\bibnamefont {Keesling}}, \bibinfo {author}
  {\bibfnamefont {G.}~\bibnamefont {Semeghini}}, \bibinfo {author}
  {\bibfnamefont {A.}~\bibnamefont {Omran}}, \bibinfo {author} {\bibfnamefont
  {D.}~\bibnamefont {Bluvstein}}, \bibinfo {author} {\bibfnamefont
  {R.}~\bibnamefont {Samajdar}}, \bibinfo {author} {\bibfnamefont
  {H.}~\bibnamefont {Pichler}}, \bibinfo {author} {\bibfnamefont {W.~W.}\
  \bibnamefont {Ho}}, \emph {et~al.},\ }\bibfield  {title} {\bibinfo {title}
  {{Quantum phases of matter on a 256-atom programmable quantum simulator}},\
  }\href {https://doi.org/10.1038/s41586-021-03582-4} {\bibfield  {journal}
  {\bibinfo  {journal} {Nature}\ }\textbf {\bibinfo {volume} {595}},\ \bibinfo
  {pages} {227} (\bibinfo {year} {2021})}\BibitemShut {NoStop}%
\bibitem [{\citenamefont {Taylor}\ \emph {et~al.}(2022)\citenamefont {Taylor},
  \citenamefont {Goswami}, \citenamefont {Walther}, \citenamefont {Spanner},
  \citenamefont {Simon},\ and\ \citenamefont {Heshami}}]{taylor2022simulation}%
  \BibitemOpen
  \bibfield  {author} {\bibinfo {author} {\bibfnamefont {J.}~\bibnamefont
  {Taylor}}, \bibinfo {author} {\bibfnamefont {S.}~\bibnamefont {Goswami}},
  \bibinfo {author} {\bibfnamefont {V.}~\bibnamefont {Walther}}, \bibinfo
  {author} {\bibfnamefont {M.}~\bibnamefont {Spanner}}, \bibinfo {author}
  {\bibfnamefont {C.}~\bibnamefont {Simon}},\ and\ \bibinfo {author}
  {\bibfnamefont {K.}~\bibnamefont {Heshami}},\ }\bibfield  {title} {\bibinfo
  {title} {{Simulation of many-body dynamics using Rydberg excitons}},\ }\href
  {https://doi.org/10.1088/2058-9565/ac70f4} {\bibfield  {journal} {\bibinfo
  {journal} {Quantum Science and Technology}\ }\textbf {\bibinfo {volume}
  {7}},\ \bibinfo {pages} {035016} (\bibinfo {year} {2022})}\BibitemShut
  {NoStop}%
\bibitem [{\citenamefont {Ebadi}\ \emph {et~al.}(2022)\citenamefont {Ebadi},
  \citenamefont {Keesling}, \citenamefont {Cain}, \citenamefont {Wang},
  \citenamefont {Levine}, \citenamefont {Bluvstein}, \citenamefont {Semeghini},
  \citenamefont {Omran}, \citenamefont {Liu}, \citenamefont {Samajdar} \emph
  {et~al.}}]{ebadi2022quantum}%
  \BibitemOpen
  \bibfield  {author} {\bibinfo {author} {\bibfnamefont {S.}~\bibnamefont
  {Ebadi}}, \bibinfo {author} {\bibfnamefont {A.}~\bibnamefont {Keesling}},
  \bibinfo {author} {\bibfnamefont {M.}~\bibnamefont {Cain}}, \bibinfo {author}
  {\bibfnamefont {T.~T.}\ \bibnamefont {Wang}}, \bibinfo {author}
  {\bibfnamefont {H.}~\bibnamefont {Levine}}, \bibinfo {author} {\bibfnamefont
  {D.}~\bibnamefont {Bluvstein}}, \bibinfo {author} {\bibfnamefont
  {G.}~\bibnamefont {Semeghini}}, \bibinfo {author} {\bibfnamefont
  {A.}~\bibnamefont {Omran}}, \bibinfo {author} {\bibfnamefont {J.-G.}\
  \bibnamefont {Liu}}, \bibinfo {author} {\bibfnamefont {R.}~\bibnamefont
  {Samajdar}}, \emph {et~al.},\ }\bibfield  {title} {\bibinfo {title} {{Quantum
  optimization of maximum independent set using Rydberg atom arrays}},\ }\href
  {https://doi.org/10.1126/science.abo6587} {\bibfield  {journal} {\bibinfo
  {journal} {Science}\ }\textbf {\bibinfo {volume} {376}},\ \bibinfo {pages}
  {1209} (\bibinfo {year} {2022})}\BibitemShut {NoStop}%
\bibitem [{\citenamefont {Lykov}\ \emph {et~al.}(2023)\citenamefont {Lykov},
  \citenamefont {Wurtz}, \citenamefont {Poole}, \citenamefont {Saffman},
  \citenamefont {Noel},\ and\ \citenamefont {Alexeev}}]{lykov2023sampling}%
  \BibitemOpen
  \bibfield  {author} {\bibinfo {author} {\bibfnamefont {D.}~\bibnamefont
  {Lykov}}, \bibinfo {author} {\bibfnamefont {J.}~\bibnamefont {Wurtz}},
  \bibinfo {author} {\bibfnamefont {C.}~\bibnamefont {Poole}}, \bibinfo
  {author} {\bibfnamefont {M.}~\bibnamefont {Saffman}}, \bibinfo {author}
  {\bibfnamefont {T.}~\bibnamefont {Noel}},\ and\ \bibinfo {author}
  {\bibfnamefont {Y.}~\bibnamefont {Alexeev}},\ }\bibfield  {title} {\bibinfo
  {title} {{Sampling frequency thresholds for the quantum advantage of the
  quantum approximate optimization algorithm}},\ }\href
  {https://doi.org/10.1038/s41534-023-00718-4} {\bibfield  {journal} {\bibinfo
  {journal} {npj Quantum Information}\ }\textbf {\bibinfo {volume} {9}},\
  \bibinfo {pages} {73} (\bibinfo {year} {2023})}\BibitemShut {NoStop}%
\bibitem [{\citenamefont {Andrist}\ \emph {et~al.}(2023)\citenamefont
  {Andrist}, \citenamefont {Schuetz}, \citenamefont {Minssen}, \citenamefont
  {Yalovetzky}, \citenamefont {Chakrabarti}, \citenamefont {Herman},
  \citenamefont {Kumar}, \citenamefont {Salton}, \citenamefont {Shaydulin},
  \citenamefont {Sun} \emph {et~al.}}]{andrist2023hardness}%
  \BibitemOpen
  \bibfield  {author} {\bibinfo {author} {\bibfnamefont {R.~S.}\ \bibnamefont
  {Andrist}}, \bibinfo {author} {\bibfnamefont {M.~J.~A.}\ \bibnamefont
  {Schuetz}}, \bibinfo {author} {\bibfnamefont {P.}~\bibnamefont {Minssen}},
  \bibinfo {author} {\bibfnamefont {R.}~\bibnamefont {Yalovetzky}}, \bibinfo
  {author} {\bibfnamefont {S.}~\bibnamefont {Chakrabarti}}, \bibinfo {author}
  {\bibfnamefont {D.}~\bibnamefont {Herman}}, \bibinfo {author} {\bibfnamefont
  {N.}~\bibnamefont {Kumar}}, \bibinfo {author} {\bibfnamefont
  {G.}~\bibnamefont {Salton}}, \bibinfo {author} {\bibfnamefont
  {R.}~\bibnamefont {Shaydulin}}, \bibinfo {author} {\bibfnamefont
  {Y.}~\bibnamefont {Sun}}, \emph {et~al.},\ }\bibfield  {title} {\bibinfo
  {title} {{Hardness of the maximum-independent-set problem on unit-disk graphs
  and prospects for quantum speedups}},\ }\href
  {https://doi.org/10.1103/PhysRevResearch.5.043277} {\bibfield  {journal}
  {\bibinfo  {journal} {Physical Review Research}\ }\textbf {\bibinfo {volume}
  {5}},\ \bibinfo {pages} {043277} (\bibinfo {year} {2023})}\BibitemShut
  {NoStop}%
\bibitem [{\citenamefont {Coelho}\ \emph {et~al.}(2022)\citenamefont {Coelho},
  \citenamefont {D'Arcangelo},\ and\ \citenamefont
  {Henry}}]{coelho2022efficient}%
  \BibitemOpen
  \bibfield  {author} {\bibinfo {author} {\bibfnamefont {W.~d.~S.}\
  \bibnamefont {Coelho}}, \bibinfo {author} {\bibfnamefont {M.}~\bibnamefont
  {D'Arcangelo}},\ and\ \bibinfo {author} {\bibfnamefont {L.-P.}\ \bibnamefont
  {Henry}},\ }\bibfield  {title} {\bibinfo {title} {{Efficient protocol for
  solving combinatorial graph problems on neutral-atom quantum processors}},\
  }\bibfield  {journal} {\bibinfo  {journal} {arXiv preprint arXiv:2207.13030}\
  }\href {https://doi.org/10.48550/arXiv.2207.13030}
  {10.48550/arXiv.2207.13030} (\bibinfo {year} {2022})\BibitemShut {NoStop}%
\bibitem [{\citenamefont {Lu}\ \emph {et~al.}(2024)\citenamefont {Lu},
  \citenamefont {Jiao}, \citenamefont {Wolinski}, \citenamefont
  {Kornja{\v{c}}a}, \citenamefont {Hu}, \citenamefont {Cantu}, \citenamefont
  {Liu}, \citenamefont {Yelin},\ and\ \citenamefont {Wang}}]{lu2024digital}%
  \BibitemOpen
  \bibfield  {author} {\bibinfo {author} {\bibfnamefont {J.~Z.}\ \bibnamefont
  {Lu}}, \bibinfo {author} {\bibfnamefont {L.}~\bibnamefont {Jiao}}, \bibinfo
  {author} {\bibfnamefont {K.}~\bibnamefont {Wolinski}}, \bibinfo {author}
  {\bibfnamefont {M.}~\bibnamefont {Kornja{\v{c}}a}}, \bibinfo {author}
  {\bibfnamefont {H.-Y.}\ \bibnamefont {Hu}}, \bibinfo {author} {\bibfnamefont
  {S.}~\bibnamefont {Cantu}}, \bibinfo {author} {\bibfnamefont
  {F.}~\bibnamefont {Liu}}, \bibinfo {author} {\bibfnamefont {S.~F.}\
  \bibnamefont {Yelin}},\ and\ \bibinfo {author} {\bibfnamefont {S.-T.}\
  \bibnamefont {Wang}},\ }\bibfield  {title} {\bibinfo {title} {{Digital-analog
  quantum learning on Rydberg atom arrays}},\ }\bibfield  {journal} {\bibinfo
  {journal} {arXiv preprint arXiv:2401.02940}\ }\href
  {https://doi.org/10.48550/arXiv.2401.02940} {10.48550/arXiv.2401.02940}
  (\bibinfo {year} {2024})\BibitemShut {NoStop}%
\bibitem [{\citenamefont {K{\"o}se}\ and\ \citenamefont
  {M{\'e}dard}(2017)}]{kose2017scheduling}%
  \BibitemOpen
  \bibfield  {author} {\bibinfo {author} {\bibfnamefont {A.}~\bibnamefont
  {K{\"o}se}}\ and\ \bibinfo {author} {\bibfnamefont {M.}~\bibnamefont
  {M{\'e}dard}},\ }\bibfield  {title} {\bibinfo {title} {{Scheduling wireless
  ad hoc networks in polynomial time using claw-free conflict graphs}},\ }in\
  \href {https://doi.org/10.1109/PIMRC.2017.8292404} {\emph {\bibinfo
  {booktitle} {Proceedings of the 2017 IEEE 28th Annual International Symposium
  on Personal, Indoor, and Mobile Radio Communications}}}\ (\bibinfo
  {organization} {Montreal IEEE, Piscataway},\ \bibinfo {year} {2017})\ pp.\
  \bibinfo {pages} {1--7}\BibitemShut {NoStop}%
\bibitem [{\citenamefont {Fin\ifmmode~\check{z}\else \v{z}\fi{}gar}\ \emph
  {et~al.}(2024)\citenamefont {Fin\ifmmode~\check{z}\else \v{z}\fi{}gar},
  \citenamefont {Schuetz}, \citenamefont {Brubaker}, \citenamefont
  {Nishimori},\ and\ \citenamefont {Katzgraber}}]{finvzgar2023designing}%
  \BibitemOpen
  \bibfield  {author} {\bibinfo {author} {\bibfnamefont {J.~R.}\ \bibnamefont
  {Fin\ifmmode~\check{z}\else \v{z}\fi{}gar}}, \bibinfo {author} {\bibfnamefont
  {M.~J.~A.}\ \bibnamefont {Schuetz}}, \bibinfo {author} {\bibfnamefont
  {J.~K.}\ \bibnamefont {Brubaker}}, \bibinfo {author} {\bibfnamefont
  {H.}~\bibnamefont {Nishimori}},\ and\ \bibinfo {author} {\bibfnamefont
  {H.~G.}\ \bibnamefont {Katzgraber}},\ }\bibfield  {title} {\bibinfo {title}
  {{Designing quantum annealing schedules using Bayesian optimization}},\
  }\href {https://doi.org/10.1103/PhysRevResearch.6.023063} {\bibfield
  {journal} {\bibinfo  {journal} {Phys. Rev. Res.}\ }\textbf {\bibinfo {volume}
  {6}},\ \bibinfo {pages} {023063} (\bibinfo {year} {2024})}\BibitemShut
  {NoStop}%
\bibitem [{\citenamefont {Przulj}(2005)}]{przulj2005graph}%
  \BibitemOpen
  \bibfield  {author} {\bibinfo {author} {\bibfnamefont {N.}~\bibnamefont
  {Przulj}},\ }\bibfield  {title} {\bibinfo {title} {{Graph theory analysis of
  protein-protein interactions}},\ }\href@noop {} {\bibfield  {journal}
  {\bibinfo  {journal} {Knowledge Discovery in Proteomics}\ }\textbf {\bibinfo
  {volume} {8}},\ \bibinfo {pages} {73} (\bibinfo {year} {2005})}\BibitemShut
  {NoStop}%
\bibitem [{\citenamefont {Byun}\ \emph {et~al.}(2022)\citenamefont {Byun},
  \citenamefont {Kim},\ and\ \citenamefont {Ahn}}]{byun2022finding}%
  \BibitemOpen
  \bibfield  {author} {\bibinfo {author} {\bibfnamefont {A.}~\bibnamefont
  {Byun}}, \bibinfo {author} {\bibfnamefont {M.}~\bibnamefont {Kim}},\ and\
  \bibinfo {author} {\bibfnamefont {J.}~\bibnamefont {Ahn}},\ }\bibfield
  {title} {\bibinfo {title} {{Finding the maximum independent sets of platonic
  graphs using Rydberg atoms}},\ }\href
  {https://doi.org/10.1103/PRXQuantum.3.030305} {\bibfield  {journal} {\bibinfo
   {journal} {PRX Quantum}\ }\textbf {\bibinfo {volume} {3}},\ \bibinfo {pages}
  {030305} (\bibinfo {year} {2022})}\BibitemShut {NoStop}%
\bibitem [{\citenamefont {Dalyac}\ \emph {et~al.}(2023)\citenamefont {Dalyac},
  \citenamefont {Henry}, \citenamefont {Kim}, \citenamefont {Ahn},\ and\
  \citenamefont {Henriet}}]{dalyac2023exploring}%
  \BibitemOpen
  \bibfield  {author} {\bibinfo {author} {\bibfnamefont {C.}~\bibnamefont
  {Dalyac}}, \bibinfo {author} {\bibfnamefont {L.-P.}\ \bibnamefont {Henry}},
  \bibinfo {author} {\bibfnamefont {M.}~\bibnamefont {Kim}}, \bibinfo {author}
  {\bibfnamefont {J.}~\bibnamefont {Ahn}},\ and\ \bibinfo {author}
  {\bibfnamefont {L.}~\bibnamefont {Henriet}},\ }\bibfield  {title} {\bibinfo
  {title} {{Exploring the impact of graph locality for the resolution of the
  maximum-independent-set problem with neutral atom devices}},\ }\href
  {https://doi.org/10.1103/PhysRevA.108.052423} {\bibfield  {journal} {\bibinfo
   {journal} {Physical Review A}\ }\textbf {\bibinfo {volume} {108}},\ \bibinfo
  {pages} {052423} (\bibinfo {year} {2023})}\BibitemShut {NoStop}%
\bibitem [{\citenamefont {Kim}\ \emph {et~al.}(2024)\citenamefont {Kim},
  \citenamefont {Kim}, \citenamefont {Park}, \citenamefont {Byun},\ and\
  \citenamefont {Ahn}}]{kim2023quantum}%
  \BibitemOpen
  \bibfield  {author} {\bibinfo {author} {\bibfnamefont {K.}~\bibnamefont
  {Kim}}, \bibinfo {author} {\bibfnamefont {M.}~\bibnamefont {Kim}}, \bibinfo
  {author} {\bibfnamefont {J.}~\bibnamefont {Park}}, \bibinfo {author}
  {\bibfnamefont {A.}~\bibnamefont {Byun}},\ and\ \bibinfo {author}
  {\bibfnamefont {J.}~\bibnamefont {Ahn}},\ }\bibfield  {title} {\bibinfo
  {title} {{Quantum computing dataset of maximum independent set problem on
  king lattice of over hundred Rydberg atoms}},\ }\href
  {https://doi.org/10.1038/s41597-024-02926-9} {\bibfield  {journal} {\bibinfo
  {journal} {Scientific Data}\ }\textbf {\bibinfo {volume} {11}},\ \bibinfo
  {pages} {111} (\bibinfo {year} {2024})}\BibitemShut {NoStop}%
\bibitem [{\citenamefont {Schiffer}\ \emph {et~al.}(2024)\citenamefont
  {Schiffer}, \citenamefont {Wild}, \citenamefont {Maskara}, \citenamefont
  {Cain}, \citenamefont {Lukin},\ and\ \citenamefont
  {Samajdar}}]{schiffer2023circumventing}%
  \BibitemOpen
  \bibfield  {author} {\bibinfo {author} {\bibfnamefont {B.~F.}\ \bibnamefont
  {Schiffer}}, \bibinfo {author} {\bibfnamefont {D.~S.}\ \bibnamefont {Wild}},
  \bibinfo {author} {\bibfnamefont {N.}~\bibnamefont {Maskara}}, \bibinfo
  {author} {\bibfnamefont {M.}~\bibnamefont {Cain}}, \bibinfo {author}
  {\bibfnamefont {M.~D.}\ \bibnamefont {Lukin}},\ and\ \bibinfo {author}
  {\bibfnamefont {R.}~\bibnamefont {Samajdar}},\ }\bibfield  {title} {\bibinfo
  {title} {{Circumventing superexponential runtimes for hard instances of
  quantum adiabatic optimization}},\ }\href
  {https://doi.org/10.1103/PhysRevResearch.6.013271} {\bibfield  {journal}
  {\bibinfo  {journal} {Phys. Rev. Res.}\ }\textbf {\bibinfo {volume} {6}},\
  \bibinfo {pages} {013271} (\bibinfo {year} {2024})}\BibitemShut {NoStop}%
\bibitem [{\citenamefont {Goswami}\ \emph {et~al.}(2024)\citenamefont
  {Goswami}, \citenamefont {Mukherjee}, \citenamefont {Ott},\ and\
  \citenamefont {Schmelcher}}]{goswami2023solving}%
  \BibitemOpen
  \bibfield  {author} {\bibinfo {author} {\bibfnamefont {K.}~\bibnamefont
  {Goswami}}, \bibinfo {author} {\bibfnamefont {R.}~\bibnamefont {Mukherjee}},
  \bibinfo {author} {\bibfnamefont {H.}~\bibnamefont {Ott}},\ and\ \bibinfo
  {author} {\bibfnamefont {P.}~\bibnamefont {Schmelcher}},\ }\bibfield  {title}
  {\bibinfo {title} {{Solving optimization problems with local light-shift
  encoding on Rydberg quantum annealers}},\ }\href
  {https://doi.org/10.1103/PhysRevResearch.6.023031} {\bibfield  {journal}
  {\bibinfo  {journal} {Phys. Rev. Res.}\ }\textbf {\bibinfo {volume} {6}},\
  \bibinfo {pages} {023031} (\bibinfo {year} {2024})}\BibitemShut {NoStop}%
\bibitem [{\citenamefont {Yeo}\ \emph {et~al.}(2024)\citenamefont {Yeo},
  \citenamefont {Kim},\ and\ \citenamefont {Jeong}}]{yeo2024approximating}%
  \BibitemOpen
  \bibfield  {author} {\bibinfo {author} {\bibfnamefont {H.}~\bibnamefont
  {Yeo}}, \bibinfo {author} {\bibfnamefont {H.~E.}\ \bibnamefont {Kim}},\ and\
  \bibinfo {author} {\bibfnamefont {K.}~\bibnamefont {Jeong}},\ }\bibfield
  {title} {\bibinfo {title} {{Approximating maximum independent set on Rydberg
  atom arrays using local detunings}},\ }\bibfield  {journal} {\bibinfo
  {journal} {arXiv preprint arXiv:2402.09180}\ }\href
  {https://doi.org/10.48550/arXiv.2402.09180} {10.48550/arXiv.2402.09180}
  (\bibinfo {year} {2024})\BibitemShut {NoStop}%
\bibitem [{\citenamefont {Nguyen}\ \emph {et~al.}(2023)\citenamefont {Nguyen},
  \citenamefont {Liu}, \citenamefont {Wurtz}, \citenamefont {Lukin},
  \citenamefont {Wang},\ and\ \citenamefont {Pichler}}]{nguyen2023quantum}%
  \BibitemOpen
  \bibfield  {author} {\bibinfo {author} {\bibfnamefont {M.-T.}\ \bibnamefont
  {Nguyen}}, \bibinfo {author} {\bibfnamefont {J.-G.}\ \bibnamefont {Liu}},
  \bibinfo {author} {\bibfnamefont {J.}~\bibnamefont {Wurtz}}, \bibinfo
  {author} {\bibfnamefont {M.~D.}\ \bibnamefont {Lukin}}, \bibinfo {author}
  {\bibfnamefont {S.-T.}\ \bibnamefont {Wang}},\ and\ \bibinfo {author}
  {\bibfnamefont {H.}~\bibnamefont {Pichler}},\ }\bibfield  {title} {\bibinfo
  {title} {{Quantum optimization with arbitrary connectivity using Rydberg atom
  arrays}},\ }\href {https://doi.org/10.1103/PRXQuantum.4.010316} {\bibfield
  {journal} {\bibinfo  {journal} {PRX Quantum}\ }\textbf {\bibinfo {volume}
  {4}},\ \bibinfo {pages} {010316} (\bibinfo {year} {2023})}\BibitemShut
  {NoStop}%
\bibitem [{\citenamefont {Lanthaler}\ \emph {et~al.}(2023)\citenamefont
  {Lanthaler}, \citenamefont {Dlaska}, \citenamefont {Ender},\ and\
  \citenamefont {Lechner}}]{lanthaler2023rydberg}%
  \BibitemOpen
  \bibfield  {author} {\bibinfo {author} {\bibfnamefont {M.}~\bibnamefont
  {Lanthaler}}, \bibinfo {author} {\bibfnamefont {C.}~\bibnamefont {Dlaska}},
  \bibinfo {author} {\bibfnamefont {K.}~\bibnamefont {Ender}},\ and\ \bibinfo
  {author} {\bibfnamefont {W.}~\bibnamefont {Lechner}},\ }\bibfield  {title}
  {\bibinfo {title} {{Rydberg-blockade-based parity quantum optimization}},\
  }\href {https://doi.org/10.1103/PhysRevLett.130.220601} {\bibfield  {journal}
  {\bibinfo  {journal} {Physical Review Letters}\ }\textbf {\bibinfo {volume}
  {130}},\ \bibinfo {pages} {220601} (\bibinfo {year} {2023})}\BibitemShut
  {NoStop}%
\bibitem [{\citenamefont {Beterov}\ and\ \citenamefont
  {Saffman}(2015)}]{PhysRevA.92.042710}%
  \BibitemOpen
  \bibfield  {author} {\bibinfo {author} {\bibfnamefont {I.~I.}\ \bibnamefont
  {Beterov}}\ and\ \bibinfo {author} {\bibfnamefont {M.}~\bibnamefont
  {Saffman}},\ }\bibfield  {title} {\bibinfo {title} {{Rydberg blockade,
  F\"orster resonances, and quantum state measurements with different atomic
  species}},\ }\href {https://doi.org/10.1103/PhysRevA.92.042710} {\bibfield
  {journal} {\bibinfo  {journal} {Phys. Rev. A}\ }\textbf {\bibinfo {volume}
  {92}},\ \bibinfo {pages} {042710} (\bibinfo {year} {2015})}\BibitemShut
  {NoStop}%
\bibitem [{\citenamefont {Hen}\ and\ \citenamefont
  {Sarandy}(2016)}]{PhysRevA.93.062312}%
  \BibitemOpen
  \bibfield  {author} {\bibinfo {author} {\bibfnamefont {I.}~\bibnamefont
  {Hen}}\ and\ \bibinfo {author} {\bibfnamefont {M.~S.}\ \bibnamefont
  {Sarandy}},\ }\bibfield  {title} {\bibinfo {title} {{Driver Hamiltonians for
  constrained optimization in quantum annealing}},\ }\href
  {https://doi.org/10.1103/PhysRevA.93.062312} {\bibfield  {journal} {\bibinfo
  {journal} {Phys. Rev. A}\ }\textbf {\bibinfo {volume} {93}},\ \bibinfo
  {pages} {062312} (\bibinfo {year} {2016})}\BibitemShut {NoStop}%
\bibitem [{\citenamefont {Kivlichan}\ \emph {et~al.}(2017)\citenamefont
  {Kivlichan}, \citenamefont {Wiebe}, \citenamefont {Babbush},\ and\
  \citenamefont {Aspuru-Guzik}}]{kivlichan2017bounding}%
  \BibitemOpen
  \bibfield  {author} {\bibinfo {author} {\bibfnamefont {I.~D.}\ \bibnamefont
  {Kivlichan}}, \bibinfo {author} {\bibfnamefont {N.}~\bibnamefont {Wiebe}},
  \bibinfo {author} {\bibfnamefont {R.}~\bibnamefont {Babbush}},\ and\ \bibinfo
  {author} {\bibfnamefont {A.}~\bibnamefont {Aspuru-Guzik}},\ }\bibfield
  {title} {\bibinfo {title} {{Bounding the costs of quantum simulation of
  many-body physics in real space}},\ }\href
  {https://doi.org/10.1088/1751-8121/aa77b8} {\bibfield  {journal} {\bibinfo
  {journal} {Journal of Physics A: Mathematical and Theoretical}\ }\textbf
  {\bibinfo {volume} {50}},\ \bibinfo {pages} {305301} (\bibinfo {year}
  {2017})}\BibitemShut {NoStop}%
\bibitem [{\citenamefont {{\v{S}}ibali{\'c}}\ \emph {et~al.}(2017)\citenamefont
  {{\v{S}}ibali{\'c}}, \citenamefont {Pritchard}, \citenamefont {Adams},\ and\
  \citenamefont {Weatherill}}]{vsibalic2017arc}%
  \BibitemOpen
  \bibfield  {author} {\bibinfo {author} {\bibfnamefont {N.}~\bibnamefont
  {{\v{S}}ibali{\'c}}}, \bibinfo {author} {\bibfnamefont {J.~D.}\ \bibnamefont
  {Pritchard}}, \bibinfo {author} {\bibfnamefont {C.~S.}\ \bibnamefont
  {Adams}},\ and\ \bibinfo {author} {\bibfnamefont {K.~J.}\ \bibnamefont
  {Weatherill}},\ }\bibfield  {title} {\bibinfo {title} {{ARC: An open-source
  library for calculating properties of alkali Rydberg atoms}},\ }\href
  {https://doi.org/10.1016/j.cpc.2017.06.015} {\bibfield  {journal} {\bibinfo
  {journal} {Computer Physics Communications}\ }\textbf {\bibinfo {volume}
  {220}},\ \bibinfo {pages} {319} (\bibinfo {year} {2017})}\BibitemShut
  {NoStop}%
\bibitem [{\citenamefont {Beterov}\ \emph {et~al.}(2016)\citenamefont
  {Beterov}, \citenamefont {Saffman}, \citenamefont {Yakshina}, \citenamefont
  {Tretyakov}, \citenamefont {Entin}, \citenamefont {Bergamini}, \citenamefont
  {Kuznetsova},\ and\ \citenamefont {Ryabtsev}}]{PhysRevA.94.062307}%
  \BibitemOpen
  \bibfield  {author} {\bibinfo {author} {\bibfnamefont {I.~I.}\ \bibnamefont
  {Beterov}}, \bibinfo {author} {\bibfnamefont {M.}~\bibnamefont {Saffman}},
  \bibinfo {author} {\bibfnamefont {E.~A.}\ \bibnamefont {Yakshina}}, \bibinfo
  {author} {\bibfnamefont {D.~B.}\ \bibnamefont {Tretyakov}}, \bibinfo {author}
  {\bibfnamefont {V.~M.}\ \bibnamefont {Entin}}, \bibinfo {author}
  {\bibfnamefont {S.}~\bibnamefont {Bergamini}}, \bibinfo {author}
  {\bibfnamefont {E.~A.}\ \bibnamefont {Kuznetsova}},\ and\ \bibinfo {author}
  {\bibfnamefont {I.~I.}\ \bibnamefont {Ryabtsev}},\ }\bibfield  {title}
  {\bibinfo {title} {{Two-qubit gates using adiabatic passage of the
  Stark-tuned F\"orster resonances in Rydberg atoms}},\ }\href
  {https://doi.org/10.1103/PhysRevA.94.062307} {\bibfield  {journal} {\bibinfo
  {journal} {Phys. Rev. A}\ }\textbf {\bibinfo {volume} {94}},\ \bibinfo
  {pages} {062307} (\bibinfo {year} {2016})}\BibitemShut {NoStop}%
\bibitem [{\citenamefont {Mc~Keever}\ and\ \citenamefont
  {Lubasch}(2024)}]{PRXQuantum.5.020362}%
  \BibitemOpen
  \bibfield  {author} {\bibinfo {author} {\bibfnamefont {C.}~\bibnamefont
  {Mc~Keever}}\ and\ \bibinfo {author} {\bibfnamefont {M.}~\bibnamefont
  {Lubasch}},\ }\bibfield  {title} {\bibinfo {title} {{Towards Adiabatic
  Quantum Computing Using Compressed Quantum Circuits}},\ }\href
  {https://doi.org/10.1103/PRXQuantum.5.020362} {\bibfield  {journal} {\bibinfo
   {journal} {PRX Quantum}\ }\textbf {\bibinfo {volume} {5}},\ \bibinfo {pages}
  {020362} (\bibinfo {year} {2024})}\BibitemShut {NoStop}%
\bibitem [{\citenamefont {Browaeys}\ and\ \citenamefont
  {Lahaye}(2020)}]{browaeys2020many}%
  \BibitemOpen
  \bibfield  {author} {\bibinfo {author} {\bibfnamefont {A.}~\bibnamefont
  {Browaeys}}\ and\ \bibinfo {author} {\bibfnamefont {T.}~\bibnamefont
  {Lahaye}},\ }\bibfield  {title} {\bibinfo {title} {{Many-body physics with
  individually controlled Rydberg atoms}},\ }\href
  {https://doi.org/10.1038/s41567-019-0733-z} {\bibfield  {journal} {\bibinfo
  {journal} {Nature Physics}\ }\textbf {\bibinfo {volume} {16}},\ \bibinfo
  {pages} {132} (\bibinfo {year} {2020})}\BibitemShut {NoStop}%
\bibitem [{\citenamefont {Leroy}\ and\ \citenamefont
  {Bernstein}(1970)}]{leroy1970dissociation}%
  \BibitemOpen
  \bibfield  {author} {\bibinfo {author} {\bibfnamefont {R.~J.}\ \bibnamefont
  {Leroy}}\ and\ \bibinfo {author} {\bibfnamefont {R.~B.}\ \bibnamefont
  {Bernstein}},\ }\bibfield  {title} {\bibinfo {title} {{Dissociation energies
  of diatomic moleculles from vibrational spacings of higher levels:
  Application to the halogens}},\ }\href
  {https://doi.org/10.1016/0009-2614(70)80125-7} {\bibfield  {journal}
  {\bibinfo  {journal} {Chemical Physics Letters}\ }\textbf {\bibinfo {volume}
  {5}},\ \bibinfo {pages} {42} (\bibinfo {year} {1970})}\BibitemShut {NoStop}%
\bibitem [{\citenamefont {M.~Farouk}\ \emph {et~al.}(2023)\citenamefont
  {M.~Farouk}, \citenamefont {Beterov}, \citenamefont {Xu}, \citenamefont
  {Bergamini},\ and\ \citenamefont {Ryabtsev}}]{Farouk2023parallel}%
  \BibitemOpen
  \bibfield  {author} {\bibinfo {author} {\bibfnamefont {A.}~\bibnamefont
  {M.~Farouk}}, \bibinfo {author} {\bibfnamefont {I.~I.}\ \bibnamefont
  {Beterov}}, \bibinfo {author} {\bibfnamefont {P.}~\bibnamefont {Xu}},
  \bibinfo {author} {\bibfnamefont {S.}~\bibnamefont {Bergamini}},\ and\
  \bibinfo {author} {\bibfnamefont {I.~I.}\ \bibnamefont {Ryabtsev}},\
  }\bibfield  {title} {\bibinfo {title} {{Parallel implementation of
  CNOT$^{\text{N}}$ and C$_2$NOT$^2$ gates via homonuclear and heteronuclear
  F{\"o}rster interactions of Rydberg atoms}},\ }\href
  {https://doi.org/10.3390/photonics10111280} {\bibfield  {journal} {\bibinfo
  {journal} {Photonics}\ }\textbf {\bibinfo {volume} {10}},\ \bibinfo {pages}
  {1280} (\bibinfo {year} {2023})}\BibitemShut {NoStop}%
\bibitem [{\citenamefont {Bernien}\ \emph {et~al.}(2017)\citenamefont
  {Bernien}, \citenamefont {Schwartz}, \citenamefont {Keesling}, \citenamefont
  {Levine}, \citenamefont {Omran}, \citenamefont {Pichler}, \citenamefont
  {Choi}, \citenamefont {Zibrov}, \citenamefont {Endres}, \citenamefont
  {Greiner} \emph {et~al.}}]{bernien2017probing}%
  \BibitemOpen
  \bibfield  {author} {\bibinfo {author} {\bibfnamefont {H.}~\bibnamefont
  {Bernien}}, \bibinfo {author} {\bibfnamefont {S.}~\bibnamefont {Schwartz}},
  \bibinfo {author} {\bibfnamefont {A.}~\bibnamefont {Keesling}}, \bibinfo
  {author} {\bibfnamefont {H.}~\bibnamefont {Levine}}, \bibinfo {author}
  {\bibfnamefont {A.}~\bibnamefont {Omran}}, \bibinfo {author} {\bibfnamefont
  {H.}~\bibnamefont {Pichler}}, \bibinfo {author} {\bibfnamefont
  {S.}~\bibnamefont {Choi}}, \bibinfo {author} {\bibfnamefont {A.~S.}\
  \bibnamefont {Zibrov}}, \bibinfo {author} {\bibfnamefont {M.}~\bibnamefont
  {Endres}}, \bibinfo {author} {\bibfnamefont {M.}~\bibnamefont {Greiner}},
  \emph {et~al.},\ }\bibfield  {title} {\bibinfo {title} {{Probing many-body
  dynamics on a 51-atom quantum simulator}},\ }\href
  {https://doi.org/10.1038/nature24622} {\bibfield  {journal} {\bibinfo
  {journal} {Nature}\ }\textbf {\bibinfo {volume} {551}},\ \bibinfo {pages}
  {579} (\bibinfo {year} {2017})}\BibitemShut {NoStop}%
\bibitem [{\citenamefont {Keesling}\ \emph {et~al.}(2019)\citenamefont
  {Keesling}, \citenamefont {Omran}, \citenamefont {Levine}, \citenamefont
  {Bernien}, \citenamefont {Pichler}, \citenamefont {Choi}, \citenamefont
  {Samajdar}, \citenamefont {Schwartz}, \citenamefont {Silvi}, \citenamefont
  {Sachdev} \emph {et~al.}}]{keesling2019quantum}%
  \BibitemOpen
  \bibfield  {author} {\bibinfo {author} {\bibfnamefont {A.}~\bibnamefont
  {Keesling}}, \bibinfo {author} {\bibfnamefont {A.}~\bibnamefont {Omran}},
  \bibinfo {author} {\bibfnamefont {H.}~\bibnamefont {Levine}}, \bibinfo
  {author} {\bibfnamefont {H.}~\bibnamefont {Bernien}}, \bibinfo {author}
  {\bibfnamefont {H.}~\bibnamefont {Pichler}}, \bibinfo {author} {\bibfnamefont
  {S.}~\bibnamefont {Choi}}, \bibinfo {author} {\bibfnamefont {R.}~\bibnamefont
  {Samajdar}}, \bibinfo {author} {\bibfnamefont {S.}~\bibnamefont {Schwartz}},
  \bibinfo {author} {\bibfnamefont {P.}~\bibnamefont {Silvi}}, \bibinfo
  {author} {\bibfnamefont {S.}~\bibnamefont {Sachdev}}, \emph {et~al.},\
  }\bibfield  {title} {\bibinfo {title} {{Quantum Kibble--Zurek mechanism and
  critical dynamics on a programmable Rydberg simulator}},\ }\href
  {https://doi.org/10.1038/s41586-019-1070-1} {\bibfield  {journal} {\bibinfo
  {journal} {Nature}\ }\textbf {\bibinfo {volume} {568}},\ \bibinfo {pages}
  {207} (\bibinfo {year} {2019})}\BibitemShut {NoStop}%
\bibitem [{\citenamefont {Fang}\ \emph {et~al.}(2024)\citenamefont {Fang},
  \citenamefont {Wang}, \citenamefont {Liu}, \citenamefont {Wang},
  \citenamefont {Cimmino}, \citenamefont {Wei}, \citenamefont {Bintz},
  \citenamefont {Parr}, \citenamefont {Kemp}, \citenamefont {Ni} \emph
  {et~al.}}]{fang2024probing}%
  \BibitemOpen
  \bibfield  {author} {\bibinfo {author} {\bibfnamefont {F.}~\bibnamefont
  {Fang}}, \bibinfo {author} {\bibfnamefont {K.}~\bibnamefont {Wang}}, \bibinfo
  {author} {\bibfnamefont {V.~S.}\ \bibnamefont {Liu}}, \bibinfo {author}
  {\bibfnamefont {Y.}~\bibnamefont {Wang}}, \bibinfo {author} {\bibfnamefont
  {R.}~\bibnamefont {Cimmino}}, \bibinfo {author} {\bibfnamefont
  {J.}~\bibnamefont {Wei}}, \bibinfo {author} {\bibfnamefont {M.}~\bibnamefont
  {Bintz}}, \bibinfo {author} {\bibfnamefont {A.}~\bibnamefont {Parr}},
  \bibinfo {author} {\bibfnamefont {J.}~\bibnamefont {Kemp}}, \bibinfo {author}
  {\bibfnamefont {K.-K.}\ \bibnamefont {Ni}}, \emph {et~al.},\ }\bibfield
  {title} {\bibinfo {title} {{Probing critical phenomena in open quantum
  systems using atom arrays}},\ }\bibfield  {journal} {\bibinfo  {journal}
  {arXiv preprint arXiv:2402.15376}\ }\href
  {https://doi.org/10.48550/arXiv.2402.15376} {10.48550/arXiv.2402.15376}
  (\bibinfo {year} {2024})\BibitemShut {NoStop}%
\bibitem [{\citenamefont {Kibble}(1976)}]{kibble1976topology}%
  \BibitemOpen
  \bibfield  {author} {\bibinfo {author} {\bibfnamefont {T.~W.}\ \bibnamefont
  {Kibble}},\ }\bibfield  {title} {\bibinfo {title} {{Topology of cosmic
  domains and strings}},\ }\href {https://doi.org/10.1088/0305-4470/9/8/029}
  {\bibfield  {journal} {\bibinfo  {journal} {Journal of Physics A:
  Mathematical and General}\ }\textbf {\bibinfo {volume} {9}},\ \bibinfo
  {pages} {1387} (\bibinfo {year} {1976})}\BibitemShut {NoStop}%
\bibitem [{\citenamefont {Zurek}(1985)}]{zurek1985cosmological}%
  \BibitemOpen
  \bibfield  {author} {\bibinfo {author} {\bibfnamefont {W.~H.}\ \bibnamefont
  {Zurek}},\ }\bibfield  {title} {\bibinfo {title} {{Cosmological experiments
  in superfluid helium?}},\ }\href {https://doi.org/10.1038/317505a0}
  {\bibfield  {journal} {\bibinfo  {journal} {Nature}\ }\textbf {\bibinfo
  {volume} {317}},\ \bibinfo {pages} {505} (\bibinfo {year}
  {1985})}\BibitemShut {NoStop}%
\bibitem [{\citenamefont {Anquez}\ \emph {et~al.}(2016)\citenamefont {Anquez},
  \citenamefont {Robbins}, \citenamefont {Bharath}, \citenamefont
  {Boguslawski}, \citenamefont {Hoang},\ and\ \citenamefont
  {Chapman}}]{PhysRevLett.116.155301}%
  \BibitemOpen
  \bibfield  {author} {\bibinfo {author} {\bibfnamefont {M.}~\bibnamefont
  {Anquez}}, \bibinfo {author} {\bibfnamefont {B.~A.}\ \bibnamefont {Robbins}},
  \bibinfo {author} {\bibfnamefont {H.~M.}\ \bibnamefont {Bharath}}, \bibinfo
  {author} {\bibfnamefont {M.}~\bibnamefont {Boguslawski}}, \bibinfo {author}
  {\bibfnamefont {T.~M.}\ \bibnamefont {Hoang}},\ and\ \bibinfo {author}
  {\bibfnamefont {M.~S.}\ \bibnamefont {Chapman}},\ }\bibfield  {title}
  {\bibinfo {title} {{Quantum Kibble-Zurek Mechanism in a Spin-1 Bose-Einstein
  Condensate}},\ }\href {https://doi.org/10.1103/PhysRevLett.116.155301}
  {\bibfield  {journal} {\bibinfo  {journal} {Phys. Rev. Lett.}\ }\textbf
  {\bibinfo {volume} {116}},\ \bibinfo {pages} {155301} (\bibinfo {year}
  {2016})}\BibitemShut {NoStop}%
\bibitem [{\citenamefont {Semeghini}\ \emph {et~al.}(2021)\citenamefont
  {Semeghini}, \citenamefont {Levine}, \citenamefont {Keesling}, \citenamefont
  {Ebadi}, \citenamefont {Wang}, \citenamefont {Bluvstein}, \citenamefont
  {Verresen}, \citenamefont {Pichler}, \citenamefont {Kalinowski},
  \citenamefont {Samajdar} \emph {et~al.}}]{semeghini2021probing}%
  \BibitemOpen
  \bibfield  {author} {\bibinfo {author} {\bibfnamefont {G.}~\bibnamefont
  {Semeghini}}, \bibinfo {author} {\bibfnamefont {H.}~\bibnamefont {Levine}},
  \bibinfo {author} {\bibfnamefont {A.}~\bibnamefont {Keesling}}, \bibinfo
  {author} {\bibfnamefont {S.}~\bibnamefont {Ebadi}}, \bibinfo {author}
  {\bibfnamefont {T.~T.}\ \bibnamefont {Wang}}, \bibinfo {author}
  {\bibfnamefont {D.}~\bibnamefont {Bluvstein}}, \bibinfo {author}
  {\bibfnamefont {R.}~\bibnamefont {Verresen}}, \bibinfo {author}
  {\bibfnamefont {H.}~\bibnamefont {Pichler}}, \bibinfo {author} {\bibfnamefont
  {M.}~\bibnamefont {Kalinowski}}, \bibinfo {author} {\bibfnamefont
  {R.}~\bibnamefont {Samajdar}}, \emph {et~al.},\ }\bibfield  {title} {\bibinfo
  {title} {{Probing topological spin liquids on a programmable quantum
  simulator}},\ }\href {https://doi.org/10.1126/science.abi8794} {\bibfield
  {journal} {\bibinfo  {journal} {Science}\ }\textbf {\bibinfo {volume}
  {374}},\ \bibinfo {pages} {1242} (\bibinfo {year} {2021})}\BibitemShut
  {NoStop}%
\bibitem [{\citenamefont {Zurek}\ \emph {et~al.}(2005)\citenamefont {Zurek},
  \citenamefont {Dorner},\ and\ \citenamefont
  {Zoller}}]{PhysRevLett.95.105701}%
  \BibitemOpen
  \bibfield  {author} {\bibinfo {author} {\bibfnamefont {W.~H.}\ \bibnamefont
  {Zurek}}, \bibinfo {author} {\bibfnamefont {U.}~\bibnamefont {Dorner}},\ and\
  \bibinfo {author} {\bibfnamefont {P.}~\bibnamefont {Zoller}},\ }\bibfield
  {title} {\bibinfo {title} {{Dynamics of a Quantum Phase Transition}},\ }\href
  {https://doi.org/10.1103/PhysRevLett.95.105701} {\bibfield  {journal}
  {\bibinfo  {journal} {Phys. Rev. Lett.}\ }\textbf {\bibinfo {volume} {95}},\
  \bibinfo {pages} {105701} (\bibinfo {year} {2005})}\BibitemShut {NoStop}%
\bibitem [{\citenamefont {Huse}\ and\ \citenamefont
  {Fisher}(1984)}]{PhysRevB.29.239}%
  \BibitemOpen
  \bibfield  {author} {\bibinfo {author} {\bibfnamefont {D.~A.}\ \bibnamefont
  {Huse}}\ and\ \bibinfo {author} {\bibfnamefont {M.~E.}\ \bibnamefont
  {Fisher}},\ }\bibfield  {title} {\bibinfo {title} {{Commensurate melting,
  domain walls, and dislocations}},\ }\href
  {https://doi.org/10.1103/PhysRevB.29.239} {\bibfield  {journal} {\bibinfo
  {journal} {Phys. Rev. B}\ }\textbf {\bibinfo {volume} {29}},\ \bibinfo
  {pages} {239} (\bibinfo {year} {1984})}\BibitemShut {NoStop}%
\bibitem [{\citenamefont {Rader}\ and\ \citenamefont
  {L{\"a}uchli}(2019)}]{rader2019floating}%
  \BibitemOpen
  \bibfield  {author} {\bibinfo {author} {\bibfnamefont {M.}~\bibnamefont
  {Rader}}\ and\ \bibinfo {author} {\bibfnamefont {A.~M.}\ \bibnamefont
  {L{\"a}uchli}},\ }\bibfield  {title} {\bibinfo {title} {{Floating phases in
  one-dimensional Rydberg Ising chains}},\ }\bibfield  {journal} {\bibinfo
  {journal} {arXiv preprint arXiv:1908.02068}\ }\href
  {https://doi.org/10.48550/arXiv.1908.02068} {10.48550/arXiv.1908.02068}
  (\bibinfo {year} {2019})\BibitemShut {NoStop}%
\bibitem [{\citenamefont {Maceira}\ \emph {et~al.}(2022)\citenamefont
  {Maceira}, \citenamefont {Chepiga},\ and\ \citenamefont
  {Mila}}]{PhysRevResearch.4.043102}%
  \BibitemOpen
  \bibfield  {author} {\bibinfo {author} {\bibfnamefont {I.~A.}\ \bibnamefont
  {Maceira}}, \bibinfo {author} {\bibfnamefont {N.}~\bibnamefont {Chepiga}},\
  and\ \bibinfo {author} {\bibfnamefont {F.}~\bibnamefont {Mila}},\ }\bibfield
  {title} {\bibinfo {title} {{Conformal and chiral phase transitions in Rydberg
  chains}},\ }\href {https://doi.org/10.1103/PhysRevResearch.4.043102}
  {\bibfield  {journal} {\bibinfo  {journal} {Phys. Rev. Res.}\ }\textbf
  {\bibinfo {volume} {4}},\ \bibinfo {pages} {043102} (\bibinfo {year}
  {2022})}\BibitemShut {NoStop}%
\bibitem [{\citenamefont {Zhang}\ \emph {et~al.}(2024)\citenamefont {Zhang},
  \citenamefont {Cant{\'u}}, \citenamefont {Liu}, \citenamefont {Bylinskii},
  \citenamefont {Braverman}, \citenamefont {Huber}, \citenamefont
  {Amato-Grill}, \citenamefont {Lukin}, \citenamefont {Gemelke}, \citenamefont
  {Keesling} \emph {et~al.}}]{zhang2024probing}%
  \BibitemOpen
  \bibfield  {author} {\bibinfo {author} {\bibfnamefont {J.}~\bibnamefont
  {Zhang}}, \bibinfo {author} {\bibfnamefont {S.~H.}\ \bibnamefont
  {Cant{\'u}}}, \bibinfo {author} {\bibfnamefont {F.}~\bibnamefont {Liu}},
  \bibinfo {author} {\bibfnamefont {A.}~\bibnamefont {Bylinskii}}, \bibinfo
  {author} {\bibfnamefont {B.}~\bibnamefont {Braverman}}, \bibinfo {author}
  {\bibfnamefont {F.}~\bibnamefont {Huber}}, \bibinfo {author} {\bibfnamefont
  {J.}~\bibnamefont {Amato-Grill}}, \bibinfo {author} {\bibfnamefont
  {A.}~\bibnamefont {Lukin}}, \bibinfo {author} {\bibfnamefont
  {N.}~\bibnamefont {Gemelke}}, \bibinfo {author} {\bibfnamefont
  {A.}~\bibnamefont {Keesling}}, \emph {et~al.},\ }\bibfield  {title} {\bibinfo
  {title} {{Probing quantum floating phases in Rydberg atom arrays}},\
  }\bibfield  {journal} {\bibinfo  {journal} {arXiv preprint arXiv:2401.08087}\
  }\href {https://doi.org/10.48550/arXiv.2401.08087}
  {10.48550/arXiv.2401.08087} (\bibinfo {year} {2024})\BibitemShut {NoStop}%
\bibitem [{\citenamefont {Balewski}\ \emph {et~al.}(2024)\citenamefont
  {Balewski}, \citenamefont {Kornjaca}, \citenamefont {Klymko}, \citenamefont
  {Darbha}, \citenamefont {Hirsbrunner}, \citenamefont {Lopes}, \citenamefont
  {Liu},\ and\ \citenamefont {Camps}}]{balewski2024engineering}%
  \BibitemOpen
  \bibfield  {author} {\bibinfo {author} {\bibfnamefont {J.}~\bibnamefont
  {Balewski}}, \bibinfo {author} {\bibfnamefont {M.}~\bibnamefont {Kornjaca}},
  \bibinfo {author} {\bibfnamefont {K.}~\bibnamefont {Klymko}}, \bibinfo
  {author} {\bibfnamefont {S.}~\bibnamefont {Darbha}}, \bibinfo {author}
  {\bibfnamefont {M.~R.}\ \bibnamefont {Hirsbrunner}}, \bibinfo {author}
  {\bibfnamefont {P.}~\bibnamefont {Lopes}}, \bibinfo {author} {\bibfnamefont
  {F.}~\bibnamefont {Liu}},\ and\ \bibinfo {author} {\bibfnamefont
  {D.}~\bibnamefont {Camps}},\ }\bibfield  {title} {\bibinfo {title}
  {{Engineering quantum states with neutral atoms}},\ }\bibfield  {journal}
  {\bibinfo  {journal} {arXiv preprint arXiv:2404.04411}\ }\href
  {https://doi.org/10.48550/arXiv.2404.04411} {10.48550/arXiv.2404.04411}
  (\bibinfo {year} {2024})\BibitemShut {NoStop}%
\bibitem [{\citenamefont {Fendley}\ \emph {et~al.}(2004)\citenamefont
  {Fendley}, \citenamefont {Sengupta},\ and\ \citenamefont
  {Sachdev}}]{PhysRevB.69.075106}%
  \BibitemOpen
  \bibfield  {author} {\bibinfo {author} {\bibfnamefont {P.}~\bibnamefont
  {Fendley}}, \bibinfo {author} {\bibfnamefont {K.}~\bibnamefont {Sengupta}},\
  and\ \bibinfo {author} {\bibfnamefont {S.}~\bibnamefont {Sachdev}},\
  }\bibfield  {title} {\bibinfo {title} {{Competing density-wave orders in a
  one-dimensional hard-boson model}},\ }\href
  {https://doi.org/10.1103/PhysRevB.69.075106} {\bibfield  {journal} {\bibinfo
  {journal} {Phys. Rev. B}\ }\textbf {\bibinfo {volume} {69}},\ \bibinfo
  {pages} {075106} (\bibinfo {year} {2004})}\BibitemShut {NoStop}%
\bibitem [{\citenamefont {Samajdar}\ \emph {et~al.}(2018)\citenamefont
  {Samajdar}, \citenamefont {Choi}, \citenamefont {Pichler}, \citenamefont
  {Lukin},\ and\ \citenamefont {Sachdev}}]{PhysRevA.98.023614}%
  \BibitemOpen
  \bibfield  {author} {\bibinfo {author} {\bibfnamefont {R.}~\bibnamefont
  {Samajdar}}, \bibinfo {author} {\bibfnamefont {S.}~\bibnamefont {Choi}},
  \bibinfo {author} {\bibfnamefont {H.}~\bibnamefont {Pichler}}, \bibinfo
  {author} {\bibfnamefont {M.~D.}\ \bibnamefont {Lukin}},\ and\ \bibinfo
  {author} {\bibfnamefont {S.}~\bibnamefont {Sachdev}},\ }\bibfield  {title}
  {\bibinfo {title} {{Numerical study of the chiral ${\mathbb{Z}}_{3}$ quantum
  phase transition in one spatial dimension}},\ }\href
  {https://doi.org/10.1103/PhysRevA.98.023614} {\bibfield  {journal} {\bibinfo
  {journal} {Phys. Rev. A}\ }\textbf {\bibinfo {volume} {98}},\ \bibinfo
  {pages} {023614} (\bibinfo {year} {2018})}\BibitemShut {NoStop}%
\bibitem [{Note1()}]{Note1}%
  \BibitemOpen
  \bibinfo {note} {We consider here the numerical value of $\protect \mathcal
  {C}_{6}$.}\BibitemShut {Stop}%
\bibitem [{\citenamefont {Byun}\ \emph {et~al.}(2024)\citenamefont {Byun},
  \citenamefont {Jung}, \citenamefont {Kim}, \citenamefont {Kim}, \citenamefont
  {Jeong}, \citenamefont {Jeong},\ and\ \citenamefont {Ahn}}]{byun2023rydberg}%
  \BibitemOpen
  \bibfield  {author} {\bibinfo {author} {\bibfnamefont {A.}~\bibnamefont
  {Byun}}, \bibinfo {author} {\bibfnamefont {J.}~\bibnamefont {Jung}}, \bibinfo
  {author} {\bibfnamefont {K.}~\bibnamefont {Kim}}, \bibinfo {author}
  {\bibfnamefont {M.}~\bibnamefont {Kim}}, \bibinfo {author} {\bibfnamefont
  {S.}~\bibnamefont {Jeong}}, \bibinfo {author} {\bibfnamefont
  {H.}~\bibnamefont {Jeong}},\ and\ \bibinfo {author} {\bibfnamefont
  {J.}~\bibnamefont {Ahn}},\ }\bibfield  {title} {\bibinfo {title}
  {Rydberg-atom graphs for quadratic unconstrained binary optimization
  problems},\ }\href {https://doi.org/10.1002/qute.202300398} {\bibfield
  {journal} {\bibinfo  {journal} {Advanced Quantum Technologies}\ }\textbf
  {\bibinfo {volume} {7}},\ \bibinfo {pages} {2300398} (\bibinfo {year}
  {2024})}\BibitemShut {NoStop}%
\bibitem [{\citenamefont {Johnson}(1966)}]{johnson1966convex}%
  \BibitemOpen
  \bibfield  {author} {\bibinfo {author} {\bibfnamefont {N.~W.}\ \bibnamefont
  {Johnson}},\ }\bibfield  {title} {\bibinfo {title} {{Convex polyhedra with
  regular faces}},\ }\href {https://doi.org/10.4153/CJM-1966-021-8} {\bibfield
  {journal} {\bibinfo  {journal} {Canadian Journal of Mathematics}\ }\textbf
  {\bibinfo {volume} {18}},\ \bibinfo {pages} {169} (\bibinfo {year}
  {1966})}\BibitemShut {NoStop}%
\bibitem [{Note2()}]{Note2}%
  \BibitemOpen
  \bibinfo {note} {\protect \href
  {https://www.graphclasses.org/}{https://www.graphclasses.org/}}\BibitemShut
  {NoStop}%
\bibitem [{\citenamefont {Qiu}\ \emph {et~al.}(2020)\citenamefont {Qiu},
  \citenamefont {Zoller},\ and\ \citenamefont {Li}}]{PRXQuantum.1.020311}%
  \BibitemOpen
  \bibfield  {author} {\bibinfo {author} {\bibfnamefont {X.}~\bibnamefont
  {Qiu}}, \bibinfo {author} {\bibfnamefont {P.}~\bibnamefont {Zoller}},\ and\
  \bibinfo {author} {\bibfnamefont {X.}~\bibnamefont {Li}},\ }\bibfield
  {title} {\bibinfo {title} {{Programmable Quantum Annealing Architectures with
  Ising Quantum Wires}},\ }\href {https://doi.org/10.1103/PRXQuantum.1.020311}
  {\bibfield  {journal} {\bibinfo  {journal} {PRX Quantum}\ }\textbf {\bibinfo
  {volume} {1}},\ \bibinfo {pages} {020311} (\bibinfo {year}
  {2020})}\BibitemShut {NoStop}%
\bibitem [{\citenamefont {Singh}\ \emph {et~al.}(2022)\citenamefont {Singh},
  \citenamefont {Anand}, \citenamefont {Pocklington}, \citenamefont {Kemp},\
  and\ \citenamefont {Bernien}}]{singh2022dual}%
  \BibitemOpen
  \bibfield  {author} {\bibinfo {author} {\bibfnamefont {K.}~\bibnamefont
  {Singh}}, \bibinfo {author} {\bibfnamefont {S.}~\bibnamefont {Anand}},
  \bibinfo {author} {\bibfnamefont {A.}~\bibnamefont {Pocklington}}, \bibinfo
  {author} {\bibfnamefont {J.~T.}\ \bibnamefont {Kemp}},\ and\ \bibinfo
  {author} {\bibfnamefont {H.}~\bibnamefont {Bernien}},\ }\bibfield  {title}
  {\bibinfo {title} {{Dual-Element, Two-Dimensional Atom Array with
  Continuous-Mode Operation}},\ }\href
  {https://doi.org/10.1103/PhysRevX.12.011040} {\bibfield  {journal} {\bibinfo
  {journal} {Physical Review X}\ }\textbf {\bibinfo {volume} {12}},\ \bibinfo
  {pages} {011040} (\bibinfo {year} {2022})}\BibitemShut {NoStop}%
\bibitem [{\citenamefont {Anand}\ \emph {et~al.}(2024)\citenamefont {Anand},
  \citenamefont {Bradley}, \citenamefont {White}, \citenamefont {Ramesh},
  \citenamefont {Singh},\ and\ \citenamefont {Bernien}}]{anand2024dual}%
  \BibitemOpen
  \bibfield  {author} {\bibinfo {author} {\bibfnamefont {S.}~\bibnamefont
  {Anand}}, \bibinfo {author} {\bibfnamefont {C.~E.}\ \bibnamefont {Bradley}},
  \bibinfo {author} {\bibfnamefont {R.}~\bibnamefont {White}}, \bibinfo
  {author} {\bibfnamefont {V.}~\bibnamefont {Ramesh}}, \bibinfo {author}
  {\bibfnamefont {K.}~\bibnamefont {Singh}},\ and\ \bibinfo {author}
  {\bibfnamefont {H.}~\bibnamefont {Bernien}},\ }\bibfield  {title} {\bibinfo
  {title} {{A dual-species Rydberg array}},\ }\bibfield  {journal} {\bibinfo
  {journal} {arXiv preprint arXiv:2401.10325}\ }\href
  {https://doi.org/10.48550/arXiv.2401.10325} {10.48550/arXiv.2401.10325}
  (\bibinfo {year} {2024})\BibitemShut {NoStop}%
\bibitem [{\citenamefont {Farouk}\ \emph {et~al.}(2023)\citenamefont {Farouk},
  \citenamefont {Beterov}, \citenamefont {Xu},\ and\ \citenamefont
  {Ryabtsev}}]{Farouk2023IJTP}%
  \BibitemOpen
  \bibfield  {author} {\bibinfo {author} {\bibfnamefont {A.~M.}\ \bibnamefont
  {Farouk}}, \bibinfo {author} {\bibfnamefont {I.~I.}\ \bibnamefont {Beterov}},
  \bibinfo {author} {\bibfnamefont {P.}~\bibnamefont {Xu}},\ and\ \bibinfo
  {author} {\bibfnamefont {I.~I.}\ \bibnamefont {Ryabtsev}},\ }\bibfield
  {title} {\bibinfo {title} {{Scalable Heteronuclear Architecture of Neutral
  Atoms Based on EIT}},\ }\href {https://doi.org/10.1134/S1063776123080046}
  {\bibfield  {journal} {\bibinfo  {journal} {Journal of Experimental and
  Theoretical Physics}\ }\textbf {\bibinfo {volume} {137}},\ \bibinfo {pages}
  {202} (\bibinfo {year} {2023})}\BibitemShut {NoStop}%
\bibitem [{\citenamefont {Ireland}\ \emph {et~al.}(2024)\citenamefont
  {Ireland}, \citenamefont {Walker},\ and\ \citenamefont
  {Pritchard}}]{ireland2024interspecies}%
  \BibitemOpen
  \bibfield  {author} {\bibinfo {author} {\bibfnamefont {P.~M.}\ \bibnamefont
  {Ireland}}, \bibinfo {author} {\bibfnamefont {D.~M.}\ \bibnamefont
  {Walker}},\ and\ \bibinfo {author} {\bibfnamefont {J.~D.}\ \bibnamefont
  {Pritchard}},\ }\bibfield  {title} {\bibinfo {title} {{Interspecies F\"orster
  resonances for Rb-Cs Rydberg $d$-states for enhanced multi-qubit gate
  fidelities}},\ }\href {https://doi.org/10.1103/PhysRevResearch.6.013293}
  {\bibfield  {journal} {\bibinfo  {journal} {Phys. Rev. Res.}\ }\textbf
  {\bibinfo {volume} {6}},\ \bibinfo {pages} {013293} (\bibinfo {year}
  {2024})}\BibitemShut {NoStop}%
\bibitem [{\citenamefont {Cesa}\ and\ \citenamefont
  {Pichler}(2023)}]{FCesaQ2023}%
  \BibitemOpen
  \bibfield  {author} {\bibinfo {author} {\bibfnamefont {F.}~\bibnamefont
  {Cesa}}\ and\ \bibinfo {author} {\bibfnamefont {H.}~\bibnamefont {Pichler}},\
  }\bibfield  {title} {\bibinfo {title} {{Universal Quantum Computation in
  Globally Driven Rydberg Atom Arrays}},\ }\href
  {https://doi.org/10.1103/PhysRevLett.131.170601} {\bibfield  {journal}
  {\bibinfo  {journal} {Phys. Rev. Lett.}\ }\textbf {\bibinfo {volume} {131}},\
  \bibinfo {pages} {170601} (\bibinfo {year} {2023})}\BibitemShut {NoStop}%
\bibitem [{Note3()}]{Note3}%
  \BibitemOpen
  \bibinfo {note} {The Rydberg blockade radius $R_{b}$ of wire atoms is
  slightly different from the Rydberg blockade radius of graph
  atoms.}\BibitemShut {Stop}%
\bibitem [{\citenamefont {de~Oliveira}\ \emph {et~al.}(2024)\citenamefont
  {de~Oliveira}, \citenamefont {Diamond-Hitchcock}, \citenamefont {Walker},
  \citenamefont {Wells-Pestell}, \citenamefont {Pelegr{\'\i}}, \citenamefont
  {Picken}, \citenamefont {Malcolm}, \citenamefont {Daley}, \citenamefont
  {Bass},\ and\ \citenamefont {Pritchard}}]{de2024demonstration}%
  \BibitemOpen
  \bibfield  {author} {\bibinfo {author} {\bibfnamefont {A.}~\bibnamefont
  {de~Oliveira}}, \bibinfo {author} {\bibfnamefont {E.}~\bibnamefont
  {Diamond-Hitchcock}}, \bibinfo {author} {\bibfnamefont {D.}~\bibnamefont
  {Walker}}, \bibinfo {author} {\bibfnamefont {M.}~\bibnamefont
  {Wells-Pestell}}, \bibinfo {author} {\bibfnamefont {G.}~\bibnamefont
  {Pelegr{\'\i}}}, \bibinfo {author} {\bibfnamefont {C.}~\bibnamefont
  {Picken}}, \bibinfo {author} {\bibfnamefont {G.}~\bibnamefont {Malcolm}},
  \bibinfo {author} {\bibfnamefont {A.}~\bibnamefont {Daley}}, \bibinfo
  {author} {\bibfnamefont {J.}~\bibnamefont {Bass}},\ and\ \bibinfo {author}
  {\bibfnamefont {J.}~\bibnamefont {Pritchard}},\ }\bibfield  {title} {\bibinfo
  {title} {{Demonstration of weighted graph optimization on a Rydberg atom
  array using local light-shifts}},\ }\bibfield  {journal} {\bibinfo  {journal}
  {arXiv preprint arXiv:2404.02658}\ }\href
  {https://doi.org/10.48550/arXiv.2404.02658} {10.48550/arXiv.2404.02658}
  (\bibinfo {year} {2024})\BibitemShut {NoStop}%
\end{thebibliography}%

%----%----%----%----%----%----%----%----%----%----%----%----%----%----%----%----%----%----%----%----%----%----%----%----%----%----%----%----%----%----%----%----%----%----%----%----%----%----%----%----%----%----%----%----%----%----%----%----%----%----%----%----%----%----%----%----%----%----%----%----%----%----%----%----%----%----%----%----%----%----%----%----%----%----%----%----%----%----%----%----%----%----%----%----%----%----%----%----%----%----%----%----%----%----%----%----%----%----%----%----%----%----%----%----%----%----%----%----%----%----%----%----%----%----%----%----%----%----%----%----%----%----%----%----%----%----%----%----%----%----%----%----%----%----%----%----%----%----%----%----%----%----%----%----%----%----%----%----%----%----%----%----%----%----%----%----%----%----%----%----%----%----%----%----%----%----%----%----%----%----%----%----%----%----%----%----%----%----%----%----%----%----%----%----%----%----%----%----%----%----%----%----%----%----%----%----%----%----%----%----%----%----%----%----%----%----%----%----%----%----%----%----%----%----%----%----%----%----%----%----%----%----%----%----%----%----%----%----%----%----%----%----%----%----%----%----%----%----%----%----%----%----%----%----%----%----%----%----%----%----%----%----%----%----%----%----%----%----%----%----%----%----%----%----%----%----%----%----%----%----%----%----%----%----%----%----%----%----%----%----%----%----%----%----%----%----%----%----%----%----%----%----%----%----%----%----%----%----%----%----%----%----%----%----%----%----%----%----
\end{document}